Analysis and Improvement of the Hot Disk Transient Plane Source Method for Low Thermal Conductivity Materials

Qiye Zheng,[1,2] Sumanjeet Kaur,[1] Chris Dames,[1,2*] and Ravi S. Prasher[1,2*]

1 Lawrence Berkeley National Laboratory, Berkeley, California

2 Mechanical Engineering, University of California at Berkeley

Abstract

The hot disk transient plane source (TPS) method is a widely used standard technique (ISO 22007-2) for the characterization of thermal properties of materials, especially the thermal conductivity, $k$. Despite its well-established reliability for a wide variety of common materials, the hot disk TPS method is also known to suffer from a substantial systematic errors when applied to low-$k$ thermal insulation materials, because of the discrepancies between the idealized model used for data analysis and the actual heat transfer process. Here, we present a combined numerical and experimental study on the influence of the geometry of hot disk sensor on measured value of low-$k$ materials. We demonstrate that the error is strongly affected by the finite thickness and thermal mass of the sensor's insulation layer was well as the corresponding increase of the effective heater size beyond the radius of the embedded metal heater itself. We also numerically investigate the dependence of the error on the sample thermal properties, confirming that the errors are worse in low-$k$ samples. A simple correction function is also provided, which converts the apparent (erroneous) result from a standard hot disk TPS measurement to a more accurate value. A standard polyimide sensor was also optimized using both wet and dry etching to provide more accurate measurement directly. Experimentally corrected value of $k$ for Airloy® x56 aerogel and a commercial silica aerogel using the numerical correction factor derived based on the standard TPS sensor is in excellent agreement with the directly measured value from the TPS sensor using the optimized polyimide sensor. Both of methods can reduce the errors to less than 2% and 4% (within the standard deviation) as compared to around 35% and 40% error of overestimation from raw values measured with the pristine sensor. This study reveals the detailed mechanisms of the systematic error in the hot disk TPS method for low-*k* samples, and show that both the numerical correction to a pristine senor or an optimized sensor are capable of providing highly accurate value of thermal conductivity for such materials.

*To whom correspondence may be addressed. Email: cdames@berkeley.edu or rsprasher@lbl.gov

**Nomenclature**

*a* temperature coefficient of electrical resistance of the metal heater

*A* slope of the linear t

*b* ring width of each heater

*c* specific heat capacity

*C* volumetric heat capacity, $C = c\rho$

*D* conventional dimensionless function to describe the temperature response of the hot disk TPS

*d* thermal penetration depth in the sample ($= 2\sqrt{\alpha_s t}$) of a transient method

*g* acceleration due to gravity

*h* thickness of a layer

$H$ new dimensionless function to describe the temperature response of the hot disk TPS proposed in this work

$j$ integer index of ring number

$k$ thermal conductivity

$L_s$ sample height

$m$ number of the concentric rings in the heater

$n$ refractive index

$p$ dimensionless radial coordinate ($= r'/r_H$) in Eq. A6

$P_{tot}$ total power input to the heater

$P_0$ areal power density into a heater with zero thickness

$q$ dimensionless radial coordinate ($= r/r_H$) in Eq. A6-A8

$(r, \theta, z)$ cylindrical coordinates

$(r', \theta', z')$ cylindrical coordinates where $r'$ and $\theta'$ are dummy variables in integrals

$r_{eff}$ effective heater radius

$r_H$ outer radius of the last ring heater

$r_j$ outer radius of $j$th ring heater ($= (j/m)r_H$ where $j = 1, 2, \ldots m$)

$r_{Kap}$ outer radius of the Kapton insulation layer

$Ra$ Rayleigh number ($= g a_1 \Delta T_{air}(h_{air})^3/\alpha \upsilon_{air}$)

$R^2$ coefficient of determination

$R_s$ sample radius

$S$ 2D area covered by the heater at the z = 0 plane covered by the heater defined in Eq. A3 to A5

$t$ heating time, measurement time

$t_c$ time correction

$t_{min}$ minimum time of the selected or optimized time range for TPS analysis

$t_{max}$ maximum time of the selected or optimized time range for TPS analysis

$t_0$ dimensionless time $(= \dfrac{z^2 + r'^2}{4\alpha_s(t)})$

$T_0$ initial temperature of the system at $t = 0$

$T_\infty$ temperature of the environment, $T_\infty = T_0$, which is approximately the far-field temperature of the sample

$\overline{\Delta T_H(t)}$ average temperature change in the metal heater defined in Eq. 1), averaged in the $(r, \theta)$ plane in the area in covered by the heater

$\overline{\Delta T_s(t)}$ temperature change at the sample surfaces in contact with the sensor, averaged in the $(r, \theta)$ plane in the area in covered by the heater

$\overline{\Delta T_i(t)}$ temperature difference across the insulation layer of the sensor, averaged in the $(r, \theta)$ plane in the area in covered by the heater

$v$ dimensionless coordinate $(= \dfrac{r'^2}{4\alpha_s(t')})$

*V* infinite 3D space for integration

**Greek symbols**

*α* thermal diffusivity, $\alpha = k/C$

*β* dimensionless ring width of the heater ($=b/r_H$)

*β'* optical absorption coefficient

*γ* thermal expansion coefficient $\sigma$ dummy variable of integral defined in Eq. 4 as well as Eq. A1 and A2.

$\tau$ ideal dimensionless time ($= \dfrac{\sqrt{\alpha_s t}}{r_H}$ with $t_c = 0$)

$\tau_c$ dimensionless time ($= \dfrac{\sqrt{\alpha_s (t - t_c)}}{r_H}$ with $t_c > 0$)

$\tau_{cl}$ lower limit of the selected or optimized range of the dimensionless time in the TPS analysis

($= \dfrac{\sqrt{\alpha_s (t_{min} - t_c)}}{r_H}$)

$\tau_{cu}$ upper limit of the selected or optimized range of the dimensionless time in the TPS analysis

($= \dfrac{\sqrt{\alpha_s (t_{max} - t_c)}}{r_H}$)

$\rho$ density

$\upsilon$ kinematic viscosity

**Subscripts**

*air* air layer in the sample gap

*app* apparent experimental measurement result

*i* insulation layer

*Kap* Kapton layer

*Ni* Ni heater layer

*s* sample

*tot* total sensor

**Superscripts**

*in* in-plane

*out* out-of-plane

1. Introduction

The rapid increase in the demand of new materials, especially porous thermal insulation materials, for thermal management in broad ranges of engineering applications demands high-throughput experimental characterization of the thermal properties of materials, i.e. the thermal conductivity, $k$, and the heat capacity, $C$. The standard steady-state techniques such as the guarded hot plate (ASTM C177), the heat flowmeter method (ISO 8301:1991), and the "cut bar" methods (ASTM D5470)[1, 2] are often accurate when conducted carefully but requires large sample size and suffer from long measurement time. In comparison, transient methods such as the transient hot wire (ASTM C518), transient plane source,[3] the laser flash method,[4] the 3-ω method,[5, 6] and the transient thermoreflectance methods[7, 8] provide higher throughput of measurement with several advantages. They can be more easily applied to small, millimeter-sized samples with relatively short measurement time and smaller unwanted heat loss, due to the controllable thermal penetration depth, $d \approx 2\sqrt{\alpha t}$, where $\alpha$ is the thermal diffusivity ($\alpha = k/C$) and $t$ is the characteristic time of the measurement.

Among transient methods, the transient plane source (TPS) technique, especially "hot disk" variants, has been widely adopted as a convenient tool for fast characterization of thermal properties.[3] This technique, now commercially available, serves as a versatile method for the simultaneous measurement of $k$ and $\alpha$ (hence $C$) in a wide range, (which are claimed to be 0.010 W m$^{-1}$ K$^{-1}$ < $k$ < 500 W m$^{-1}$ K$^{-1}$, and $5 \times 10^{-8}$ m$^2$ s$^{-1}$ < $\alpha$ < $10^{-4}$ m$^2$ s$^{-1}$) and is compatible with samples of many different forms.[9] In addition to isotropic bulk materials,[1, 10, 11] and fluid,[12, 13] the TPS methods can be extended to study slab materials, thin film specimens (dimensions < $d$) and anisotropic materials.[14] This method uses thin metal foil sealed within dielectric insulation layer as both heater and sensor. As an absolute technique, does not require extra thermal calibration steps before measuring the sample of interest. When it comes to low-$k$ solid materials, the TPS method is more tolerant to relatively rough sample surface, which is common for many porous thermal insulation (TI) materials. It also does not need metal sensors or transducer layer coatings on the sample which is necessary in traditional 3-ω, [5, 6] or optical-based methods.[4, 7, 8] A TPS measurement is also at least an order of magnitude faster compared to steady-state methods.

The hot disk TPS is stated[9] to be suitable for materials with $k$ as low as approximately 0.010 W m$^{-1}$ K$^{-1}$, although the lowest value verified with other measurement technique is 0.027 W m$^{-1}$ K$^{-1}$.[15] More recently, it has been noticed that the measurement error of the TPS method using sensors with common design where the dielectric (e.g., polyimide and mica) electrical insulation layer thickness is >30 µm can be significant for TI materials, especially those with $k$ < 0.05 W m$^{-1}$ K$^{-1}$.[16-18] For such sample materials, it has been shown that the $k$ measured by the hot disk TPS method with a standard commercial sensor with polyimide (Kapton) as the insulation layer can be 20%-50% higher than the results from the steady-state methods.[18, 19] For aerogel samples, there is also a report on the discrepancy in the $k$ result obtained with TPS measurements using different types of sensors insulated with mica and polyimide, where the former type showed a result 54% higher than the later for the same sample.[20]

Error in the TPS measurement[20-31] comes from two sources: (1) uncertainty in the experimental data and the selection of time interval for analysis, and (2) deviation of the original idealized analytical heat transfer model[3, 32] from the practical measurement scenario. Previous works on the former aspect focused on the sensitivity of the input parameters and the data fitting procedure based on the original analytical model, but could not explain the overestimation of $k$ in TI materials.[21-26] In the latter aspect, several publications investigated the accuracy and performance of commercial TPS devices based on numerical simulations to study the effect of the sensors on the TPS measurement for bulk or thin film[20, 27-31] and the error due to thermal radiation in semitransparent sample.[17, 33] However, these researches provide no systematic investigation on the sensor geometric parameters and sample thermal properties. Although it was mentioned that the finite thermal mass of the sensor can lead to error in some low-$k$ materials,[17, 20] there is still a lack of a clear understanding and quantitatively useful solution to this problem which relates the measurement error for TI materials to the sensor parameters in a definitive way. Moreover, no research has been done to enhance the measurement accuracy by directly modifying the sensor.

Here, we present a systematic study on the error of the hot disk TPS measurement for low-$k$ materials by improving the analytical model, establishing systematic numerical finite element modeling, and experimental modification of the hot disk TPS sensor (Fig. 2(a)). We first review the analytical Green's function solution of the ideal heat transfer problem and the analysis procedure and introduce a new analytical model for the problem which overcomes the known divergence issue near $t = 0$ in the original solution.[24, 34] Then, by analyzing the $k$ and $C$ results identified from the numerically simulated hot disk TPS data following the standard,[9] we discuss the correlation between the geometric and thermal properties of the sensor and the systematic error due to the discrepancy between the analytical model and the actual heat transfer. To account for the realistic heat flow in hot disk TPS, we provide a correction function to improve the measurement accuracy for the implementation of common polyimide hot disk sensors (Kapton-5501F and Kapton-7577 from Thermtest®) is provided. This study culminates by experimentally demonstrating the improvement of the TPS measurement accuracy by tens of percent by reducing the Kapton insulating layer thickness through both dry and wet etching, which validates both the numerical error analysis and the correction function.

## 2. Principle of Operation and Theoretical Basis of the TPS Method

### 2.1 Operation Principle and the Structure of This work

The hot disk TPS techniques[3, 9] use a thin metal foil disk (~10 µm thickness, ~0.5-30 mm in diameter, depending on the sensor type), usually with a bifilar spiral pattern, as both the temperature sensor and the electrical resistive heater. The schematics of the sensor and experimental configuration are shown in Fig. 1. The metal disk is sealed between two thin sheets of polyimide (Kapton), aluminum nitride, aluminum oxide, or mica, which act as a structural support and electrical insulator. During the experiment, the hot disk sensor is sandwiched between two pieces of identical samples to be tested, and a stepwise Joule heating is produced by applying a stepwise current to the sensor which generates dynamic temperature field in the sample and the sensor. By recording the increase in the resistance of the metal sensor as a function of time, through resistance thermometry, the temperature increase in the sensor is accurately monitored. The temperature response is then analyzed to deduce the thermal properties of the sample based on the model developed for the idealized sensor with known geometry, which acts as a boundary condition to the heat conduction problem in the sample. As we shall see below, the traditional idealization of the sensor can cause substantial errors for low-*k* samples.

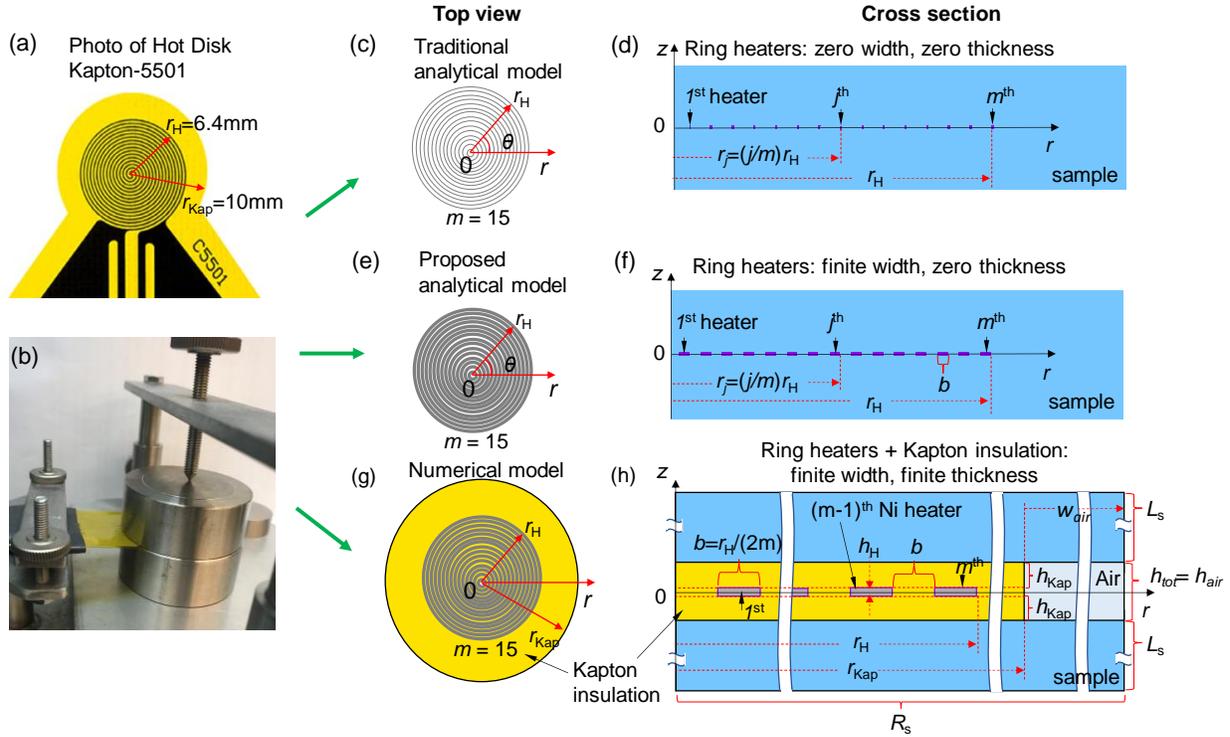

Fig. 1 Schematic of the analytical and numerical modeling for TPS measurements. (a) Top view of a real hot disk Kapton 5501 sensor with Kapton insulation. The margin of the Kapton is the region of Kapton between $r_H$ and $r_{Kap}$. (b) Photo of a hot disk TPS measurement of a stainless steel reference sample showing the configuration of the experiment, with the sensor sandwiched between the two sample halves. (c, d) Top and cross-sectional views of the hot disk TPS geometry assumed in the traditional analytical model of Eq (4). (e, f) Similar top and cross-sectional views for the new analytical model of this work. A series of m=15 ring heaters with infinitely narrow (c, d) and finite (e, f) width are labeled. (g, h) Similar top and cross-sectional views of the computational domain used in the numerical model of this work. For clarity the aspect ratio of (h) is not 1:1, and break lines are used to facilitate visualizing features across multiple scales. Various geometric parameters are labeled.

Because the temperature excursions are small, the resistance of sensor (i.e., heater) is well approximated as a linear function of the heater's average temperature rise, $\overline{\Delta T_H(t)} = \overline{T_H(t)} - T_0$ through

$$R(t) = R_0(1 + a\overline{\Delta T_H(t)}), \quad (1)$$

where $R$ is the sensor electrical resistance at time $t$, $R_0$ is the initial resistance at $t = 0$ at which point the entire system is isothermal at $T_0$, $a$ is the temperature coefficient of resistivity (TCR) calibrated prior to use for each sensor,[†] and the overbar represents spatial averaging over the space where the metal heater lies. In actual measurement and our numerical simulation where the finite thickness of the sensor is considered, $\overline{T_H(t)}$ (and $\overline{\Delta T_H(t)}$) is the average temperature (and temperature change) of the whole volume of the metal heater. By contrast, in the traditional analytical model (and the new one we propose, see below) the averaging space for $\overline{\Delta T_H(t)}$ is the length (area) covered by the metal heater in the zero-thickness sensor on the $z = 0$ plane.

We can further decompose the average temperature rise in the sensor into contributions from the sample and the sensor, by writing

$$\overline{\Delta T_H(t)} = \overline{\Delta T_s(t)} + \overline{\Delta T_i(t)} \quad (2)$$

where $\overline{\Delta T_s(t)} = \overline{T_s(t)} - T_\infty$ denotes the average temperature drop between the sample surface in contact with the sensor, $T_s(t)$, and the far field $T_\infty$ (assumed to remain at $T_0$), which is determined by the sample thermal properties. The averaging for $T_s(t)$ is performed over the projected area covered by the heater on the $z = 0$ plane. $\overline{\Delta T_i(t)} = \overline{T_H(t)} - \overline{T_s(t)}$ denotes the average temperature drop between the metal heater element and the sample-sensor interface.

---

[†] Note the influence of the resistance change of the heater on the joule heating power is a negligible, second-order effect, typically leading to errors <1% in the final results.[31]

Provided that the sensor's insulating layer is thin, under constant heating power it is well known[20] that $\overline{\Delta T_i(t)}$ becomes almost constant after a short time, a few multiples of the diffusion time $h_i^2/\alpha_i$, where $h_i$ is the thickness of the insulating layer and $\alpha_i$ is its thermal diffusivity. For a typical polyimide insulated sensor like the commercially available Kapton-5501F, with $h_i=h_{Kap}$ $\approx$25 µm (see Fig. 1), this stabilization time is less than 100 ms.[9] The temperature at the sample surface $\overline{\Delta T_s(t)}$ is approximated with an analytical model as discussed below. By analyzing $\overline{\Delta T_H(t)}$ in the time scale long enough for heat to diffuse the length of the $r_H$ in the sample after the initial diffusion through the thickness of the sensor (i.e., $t >$ 100 ms) using this thermal model, the sample $k$ and $C$ can be obtained. Although the out-of-plane heat diffusion through the thickness of the sensor quickly approaches steady state, the in-plane heat diffusion along the insulation layer in with millimeter scale of radius is slow and will significantly affect $\overline{\Delta T_H(t)}$ and the hot disk TPS measurement results as shall be discussed below.

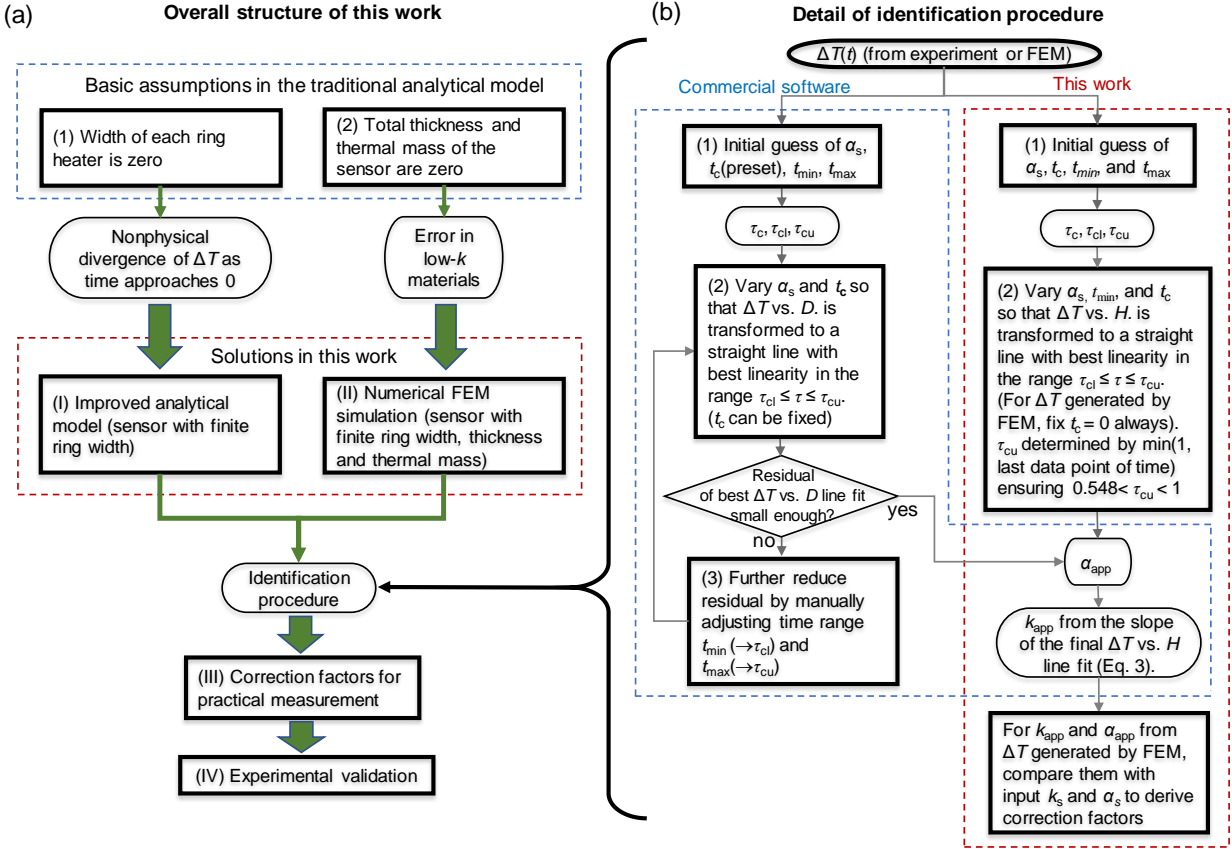

Fig. 2 (a) Structure of this work. The basic assumptions ((1) and (2)) in the traditional analytical model of the hot disk TPS measurement cause two main issues. We solve these issues by (I) developing a new analytical model and (II) using FEM simulations with more realistic geometry. (III) Simple correction factors for two common sensors are obtained using FEM simulations. (IV) Finally, these findings are validated by measurements using hot disk TPS sensors with modified geometry (thickness, radius) and comparison with separate steady-state measurements using a heat flow meter. (b) Flow charts of the identification procedure in the commercial software and this work ("2-step" procedure). The main difference between the two methods is that the manual optimization of $\tau_{cl}$ and $\tau_{cu}$ in the commercial software is merged into the automated fitting process in this work. The initial guess of $t_c$ in the commercial software is a preset small number not changeable by the user and is fixed at $t_c = 0$ in the identification procedures for data from FEM in this work. A more detailed flow chart of the identification process in this work, especially box (2), is shown in the Appendix Fig. A2.

The bifilar spiral metal heater together with the dielectric insulation layer are usually approximated by several concentric and equally spaced circular rings with constant power in both analytical[3, 32] and numerical[20, 27-31] prior models. The traditional analytical solution[3, 32] to the heat transfer problem involved in a TPS measurement of isotropic bulk materials and the corresponding analysis procedure in the standard[9] used in a common commercial software (Hot Disk Thermal Constants Analyzer®) is established based on four assumptions about geometry of the sensor and sample.[3, 32] (1) The concentric ring heaters have infinitely narrow ring width at the $z = 0$ plane. (2) The whole sensor, including the metal heater and the insulation layer, has zero total thickness (see Fig. 1(a)). (3) The sample domain dimensions are infinite. (4) The contact interface between the sample and sensor is flat and uniform. The analytical model based on the first assumption gives nonphysical divergence of temperature response of the sensor (see section 2.2). Assumption (2) is acceptable for high-$k$ materials, but as discussed later, fails for low-$k$ materials especially if the sample $\alpha$ is relatively small and lead to large systematic errors. Assumption (3) is valid if the TPS thermal penetration depth, $d = 2\sqrt{\alpha_s t}$, measured from any part of the sensor, is smaller than the sample dimensions, which is commonly satisfied for large bulk samples. Assumption (4) is also typically satisfied if the sample is flat and the measurement is done properly under vertical pressure.

In this work, we analyze the improve the hot disk TPS method by focusing on the first two assumptions with a structure described in Fig. 2(a). For issue caused by (1), an improved analytical model that assume finite heater width with zero total thickness is provided. This model better mimics the real sensor and involves no diverging term at any time. For the issue related to (2), we utilized systematic numerical simulation based on finite element modeling (FEM) to study the effect of the realistic non-ideal sensor with finite thickness and thermal mass. The virtual hot disk TPS temperature response data from such FEM is analyzed by an identification procedure developed to generate the same apparent erroneous result as the commercial software based on the standard (see flow charts in Fig. 2(b) and Fig. A2).[9] The apparent results are then compared with the input parameters to derive the relative error and correction factors for the sensor. Such correction factors relate the experimental data obtained using the commercial software to more intrinsic thermal properties of the sample.

2.2 Traditional Theory: Infinitesimal Ring Width

Based on the four assumptions mentioned above, the average temperature evolution of the heater hot disk TPS is traditionally solved using a Green's function method by integrating the instantaneous point source solution[35] and then averaging over the idealized ring sources with zero thickness and infinitely narrow width in space and time.[3, 32]

$$\overline{\Delta T_s(\tau_c)} = \frac{P_{tot}}{(\pi)^{3/2} r_H k_s} D_m(\tau_c) = A D_m(\tau_c) \tag{3}$$

where

$$D_m(\tau_c) = [m(m+1)]^{-2} \int_0^{\tau_c} \sigma^{-2} \left[ \sum_{l=1}^{m} l \sum_{k=1}^{m} k \exp\left(-\frac{l^2+k^2}{4m^2\sigma^2}\right) I_0\left(\frac{lk}{2m^2\sigma^2}\right) \right] d\sigma \tag{4}$$

and

$$\tau_c = \frac{\sqrt{(t-t_c)\alpha_s}}{r_H} \tag{5}$$

Here, $P_{tot}$ is the total constant power input at the heater, $\alpha_s$ is the sample thermal diffusivity, $k_s$ is the sample thermal conductivity, $r_H$ is the outer radius of the metal heater, $m$ is the number of the concentric rings, $A = \dfrac{P_{tot}}{(\pi)^{3/2} r_H k_s}$, and $\tau_c$ is a dimensionless characteristic time. According to the standard,[9] the time correction $t_c$, is introduced because of unavoidable hardware and software delays which cause the full power output of the sensor not to coincide exactly with the time $t = 0$. $t_c$ is hence approximately a constant and is different from the time for heat diffusion through the insulation layer $\sim h_i^2/\alpha_i$.[‡] The dummy variable of the integration $\sigma$ is defined as $\sigma = \dfrac{\sqrt{(t-t')\alpha_s}}{r_H}$, where $t'$ is the time variable in the Green's function (not shown). Combining Eq. 3 with 2, we can write

$$\overline{\Delta T_H(\tau_c)} = \overline{\Delta T_s(\tau_c)} + \overline{\Delta T_i(\tau_c)} \approx A D_m(\tau_c) + B \qquad (6)$$

for measurement time longer than $h_i^2/\alpha_i$, where $B$ is a constant to approximate $\overline{\Delta T_i(\tau_c)}$. Such linear relation between $\Delta T_H$ (referred as $\Delta T$ in Fig. 2 and A2 for simplicity) and $D_m(\tau_c)$ forms the basis of identifying procedure to obtain thermal properties in hot disk TPS measurement (Fig. 2 and A2).

The traditional model of Eq. 4,[3, 32] treats the heater as line source with zero width and thickness, and hence inevitably leads to diverging temperature in the heater as time approaches zero which can be easily shown by Taylor expanding the modified Bessel function in the integral in Eq. 4 (see Appendix section A1).[36] It is not mentioned how this issue was dealt with in the standard,[9] but in literature, a cutoff time for the lower limit of the integral in Eq. 4 was set (such that $\tau_c > 0.03$) and the result is concatenated to a 1D heat transfer approximation model at $\tau_c < 0.03$.[24] In other TPS

---

[‡] It is reasonable to assume $t_c = 0$ in the analysis of our numerically calculated temperature response data where no such delay occurs (see below). In the analysis of experimental data, we find the time corrections $t_c$ is typically less than 100 ms and is never larger than 0.1 % of the total measurement time for TI materials.

methods such as hot strip and hot square, which are also founded by Gustafsson, the finite width of the heater is considered.[3, 37]

2.3 Improved Analytical Model: Finite Ring Width

By using Green's function method, we develop an improved analytical solution to the heat transfer problem of the hot disk TPS by considering concentric ring heaters with *finite width* in infinite space which does not give without any diverging problem (see the top and cross section view of the configuration in Fig. 1 (d, e)). The purpose of this analytical model is to eliminate the mathematical issue in the original model and the uncertainty in the determination of the lower limit of the integral. The finite thickness of the sensor will be considered in our numerical model. The derivation and the complete form of the new model is included in the Appendix section A1 with the final expressions shown in Eq. A6-A11 which define a dimensionless temperature $H_m(\tau_c, \beta)$ for heater of $m$ finite-width rings ($\beta$ is the dimensionless ring width $b/r_H$). Note that $H_m(\tau_c, \beta)$ has the same relation with $\overline{\Delta T_s(\tau_c)}$ as $D_m(\tau_c)$ according to Eq. A11 and Eq. 3 and hence should have the same magnitude as $D_m(\tau_c)$ if both of them are calculated from a known correct $\overline{\Delta T_s(\tau_c)}$. Ideally, if the concatenation or other method used in fixing $D_m(\tau_c)$ works, the corrected $D_m(\tau_c)$, e.g. $D_m(\tau_c)$ shown in Ref [11], should also be the same as $H_m(\tau_c, \beta)$.

As shown in Fig. A1, the new model $H_m(\tau_c, \beta)$ agrees well with the literature result of $D_m(\tau_c)$ with $m = 15$ which is presumably obtained by the cut-off method mentioned above.[11] As expected, $D_m(\tau_c)$ from Eq. 4 strongly depends on the lower limit (lb) of integral (Fig. A1(a)) showing large shift even if the lb is changed by a tiny amount. In Fig. A1, the new analytical model results for $m = 10$ and 15 are also compared to a FEM simulation with COMSOL Multiphysics® for the TPS heat conduction problem (Fig. 1(e, f)) using sensor with finite-width and zero thickness at $z = 0$ plane in a large sample domain. The dimensionless temperature converted from the average sensor temperature from the numerical simulation with unit K (Eq. A11) agrees with the new analytical model $H_m(\tau_c, \beta)$ in Eq. A8 with the same $m$ within 0.5%.

2.4 The TPS Data Analysis Process to Identify Sample Thermal Properties

To obtain the thermal properties of a sample with unknown $\alpha_s$ and $k_s$, the TPS data is typically analyzed in three steps according to the standard[9] (see the flowcharts in Fig. 2(b) and A2).[9] (1) First, initial guesses for four quantities: $\alpha_s$, $t_c$, and the time range for fitting $[t_{min}, t_{max}]$ are made to derive the dimensionless time $\tau_c$ as well as the range of $\tau_c$ for fitting with a lower and upper bound of $\tau_{cl} = \dfrac{\sqrt{(t_{min} - t_c)\alpha_s}}{r_H}$ and $\tau_{cu} = \dfrac{\sqrt{(t_{max} - t_c)\alpha_s}}{r_H}$ respectively. (2) A linear regression of experimental data, i.e. $\overline{\Delta T_H(\tau_c)}$ vs. $D_m(\tau_c)$ (or $\overline{\Delta T_H(\tau_c)}$ vs. $H_m(\tau_c, \beta)$ in our analysis) using Eq. 6 is performed with the goodness of fitting characterized by the coefficient of determination, $R^2$ in $[\tau_{cl}, \tau_{cu}]$. Then $\alpha_s$ and $t_c$ are varied iteratively to transform the linear regression of Eq. (6) to a straight line by maximizing $R^2$ for the best linearity. (3) Finally, the time window $[t_{min}, t_{max}]$ is revisited and manually adjusted by the user, followed by further the ($\alpha_s$, $t_c$) optimization as in Step 2. This iteration between Steps 2 and 3 is repeated until the linear regression residuals are judged small enough and hence $R^2$ large enough. The apparent sample thermal diffusivity, $\alpha_{app}$, is obtained from the final step of the iteration procedure. Then, with known power input $P_{tot}$ and heater radius $r_H$, the apparent sample conductivity $k_{app}$ is determined from the slope $A$ of the final linear regression with Eq. (6). The heat capacity is derived from $C_{app} = k_{app}/\alpha_{app}$.[§]

---

[§] The intercept $B \sim \overline{\Delta T_i(\tau_c)}$ is not of interest to bulk sample measurement here.

Step (3) often changes the fit $k_{app}$ by up to ~10%, and typically can reduce the RMS fitting residual to < 1 mK, which corresponds to less than 0.1% of the total temperature rise of typically 2 - 5 K for TI materials. For example, if the initial guesses of $t_{min}$ is too small the discrepancy between the actual heat transfer and the analytical model can be significant due to the finite size of the sensor. This causes large non-random residuals in $\overline{\Delta T_H(\tau_c)}$ vs. $D_m(\tau_c)$ fitting, which deviates from a straight line and error in $\alpha_{app}$. However, the optimization (typically increasing) of $t_{min}$ aimed to eliminate the heat diffusion in the sensor at early time not accounted in the analytical model fails for TI materials due to the correlated time range of heat diffusion in the sensor and sample (see section 4). Even with the residuals < 1 mK by this additional optimization,[17] the resulting $k_{app}$ can be substantially higher than the true value of $k_s$, especially for TI materials.

According to a standard[9] and early parameter analysis,[26] $t_{max}$ should be limited in a range such that $0.3 < \tau_{cu}^2 < 1$, (or equivalently $0.548 < \tau_{cu} < 1$) to allow the TPS data to be sensitive to both sample thermal conductivity, $k_s$, and volumetric heat capacity, $C_s$. The $t_{max}$ is typically the last data point of the measurement due to the limitation of the sample size. If the measurement time is chosen to be too long such that $\tau_{cu}$ is larger than 1 (i.e. $d > 2r_H$), $t_{max}$ is simply reduced under a loose constraint to ensure $0.548 < \tau_{cu} < 1$. Hence most of the time, the user only needs to vary $t_{min}$ (equivalently, $\tau_{cl}$) to reduce the residual according to user judgment. The upper limit of $\tau_{cl}$ is set to ensure at least 5 data points within $\tau_{cl} \leq \tau_c \leq \tau_{cu}$ to maintain enough signal to noise ratio.[38]

Detailed information about the specific fitting and optimization procedure is rarely given in the literature, being typically either described simply as a least square method (often without specifying the fitting parameters)[20, 27, 28] or not clearly mentioned.[29] This details of this step were only mentioned in the early work on the TPS sensitivity and parameter estimation by Gustafsson and a few studies on a similar transient technique called dynamic plane source.[25, 26]

The commercial software Hot Disk Thermal Constants Analyzer®, embedded in the Thermtest Hot Disk TPS® is not amenable to user modification and cannot operate using synthetic data from our simulations. Therefore, a home-built program is developed to calculate the sample $k_s$ and $α_s$ from the temperature response curve of the sensor obtained by numerical simulation. This program is validated by analyzing both the exported raw experimental data of the Hot Disk TPS® in comparison with the $k_s$ and $C_s$ results from the commercial software[38] and by analyzing the numerically generated temperature data for the ideal concentric ring heaters with zero thickness (as shown in the schematic in Fig. A3). See more details in Appendix A2.

The flow chart of the identification procedure of the home-built program is shown in Fig. 2(b) and A2 in comparison with the process of the commercial software. The step (2) and (3) in the commercial software are based on the same type of optimization that maximizes $R^2$ of the linear regression (i.e. "linearity") to transform the fitting with Eq. 6 to a straight line. To avoid manual adjustment of the time range, a 2-step process that merges step (2) and (3) is used in our home-developed program. The tolerance of the optimization of 1-$R^2$ is set to $10^{-4}$ (i.e. $R^2 > 0.9999$). In the merged iteration step, 3 parameters: thermal diffusivity $α_s$, time correction $t_c$, and the lower limit of the dimensionless time, $τ_{cl}$ are adjusted to maximize $R^2$ in a least square fitting. After the final step, apparent values of $α_{app}$ and $k_{app}$ are obtained (see Fig. A2 ad Appendix A2).

3 Numerical Method and Results

3.1 Numerical Method and Computational Domains

To understand the sources of error in the TPS measurement of the low-$k$ materials, numerical modeling tool is used to generate virtual hot disk TPS data of the average sensor temperature as a function of time and then analyzed by the procedure described below. The dynamic test process of the TPS is simulated using the COMSOL Multiphysics® package. The problem is treated with a time dependent heat transfer module using the non-linear solver, MUMPS (MUltifrontal Massively Parallel Sparse direct Solver).[39]

The schematics of the computational domain consisting of the heating element (Ni), insulation layer (Kapton polyimide), and test material (sample) are plotted in Fig. 1(g, h), with most of the geometric parameters labeled. Considering the axial symmetry of the problem, the 3D heat transfer process is reduced to a 2D problem with axial symmetry geometry. As a test case we used the dimensions of the TPS sensor Kapton-5501F in Fig 1. The parameters for a Kapton-5501 (similar to 5501F with the only difference being the thickness of insulation layer) and a Kapton-7577 sensor from Thermtest® are listed in Table 1.

Fig. 1(a) shows the top view photo of a real sensor. A structure of 15 concentric rings each with width $b = r_H/30$ and finite thickness $h_{Ni} = 10$ μm where the outer radius of the outer ring $r_H = 6.403$ mm is used to approximate the real sensor with double spiral structure as shown in Fig. 1 (g, h), assuming the difference at the small center region and the four metal leads part is negligible.[**] The Ni ring heaters are sandwiched between Kapton insulation layers with outer radius of $r_{Kap} = 10$ mm and the gap between adjacent Ni rings are filled with Kapton material. The thicknesses of Kapton layer, $h_{Kap}$, are obtained based on the micrometer measurement of the real sensors assuming mirror plane symmetry at the $z = 0$ plane, and the thickness of the Ni heater $h_{Ni}$ is from the standard.[9] The total sensor thickness $h_{tot} = 2h_{Kap} + h_H$. The Kapton margin which shall be discussed later refer to the region of Kapton layer within $< r < r_{Kap}$. The edge of the Kapton sensor at $r = r_{Kap}$ is in direct contact with air. The whole sensor is sandwiched between two pieces of identical sample materials with effectively infinite height $L_s = 50$ mm and radius $R_s = 100$ mm for the TPS heat diffusion (both of which significantly larger than the TPS thermal penetration depth $d \approx 2r_H$ for all types of sensors studied, see Appendix A3). The detailed baseline thermal properties and geometric parameters of the Ni heater, the Kapton insulation layer, and the sample are listed in Table 1. Simulations are further conducted for various sensor geometric parameters and thermal properties as well as varied sample thermal properties which will be specified for different cases in later sections.

---

[**] Since the wide Ni current leads should have small resistance compared with the double spiral part and there is no current flow in the thin voltage lead, these leads are not expected to act as heaters. But the extra insulation layer covering the voltage probe and current probe outside the circular part may affect the result. This is partially taken care of in the later analysis of the systematic error from the Kapton margin width for the measurement of the low-$k$ materials.

To generate virtual hot disk TPS data, a small constant total power of $P_{tot} = 0.02$ W is applied to the Ni heaters from $t = 0$ in the simulation of the TI materials in Table 1, which leads to a maximum temperature rise of a few K in the time range of interest. For much higher-$k$ stainless steel, $P_{tot} = 2$ W was used. With temperature $\Delta T(r, z, t)$ evolution of the system, the time dependent temperature of the sensor is averaged over the whole volume of all Ni heater rings, which should be a good approximation of the experimentally determined temperature response $\overline{\Delta T_H(t)}$.[28]

3.2 Boundary Conditions and Radiation Effect on the Hot Disk TPS Measurement

The effects of inter-domain interfacial thermal resistances are (ITR) considered Appendix A6. It is shown that for ITR between the sample and sensor in the physically relevant range for common interface, it has no influence on the identified sample thermal properties. The boundary conditions (BC) at the outer surfaces of the sample and the outer wall of the air in the gap were also examined, to confirm that the finite-sized simulation domains adequately approximate the semi-infinite domains assumed in the analytical modelling. whether the far-field BC was thermally insulating or convective, with convection coefficient varied from 1 to 1000 W m$^{-2}$ K$^{-1}$ and with or without blackbody radiation, the differences in sensor temperature were always less than 0.1% throughout the entire simulation time $t < t_{max}$ ($t_{max} \sim 10^4$ s for TI samples). This confirms that the sample sizes ($R_s$ = 100 mm and thickness $L_s$ = 50 mm for each piece of sample) in our FEM simulation are large enough to be considered as infinite. We have further checked the effect of the convection of the air in the sample gap in the Appendix A3. Due the small size of the gap and hence a small Rayleigh number, the convection of air has negligible effect on heat transfer in the TPS measurement.

To study the thermal radiation on the hot disk TPS measurement, the radiation in participating medium is coupled with the heat conduction in the COMSOL simulation. We show the influence of average absorption coefficient of the sample in Appendix A6 (other simulation in the main text does not consider radiation). The key conclusion is that when the absorption coefficient of the optically thick bulk sample is larger than $2 \times 10^4$ m$^{-1}$, common for medium-density or opacified TI materials, the radiative heat transfer in the sample contribute little to the TPS result. Themal radiation was discussed in previous TPS literature[33] and the study here is for a qualitatively demonstration.

Table 1: Geometric parameters, thermal properties (*C, k* and *α*), and the optical properties (refractive index *n* and absorption coefficient *β'* at 10 um) of the materials used for the computational domains in the numerical model. The geometric parameters are based on commercial sensors of Kapton 5501, 5501F and 7577. *k* matrix of the Kapton is taken from literature.[40] Tabulated temperature dependent thermal properties of air in the COMSOL are used. The thermo-physical properties of TIA and TIC are used to mimic the experimentally measured Airloy® x56 aerogel and the reference SRM1453 polystyrene foam respectively.

A hypothetical TI material, labeled TIB, is chosen to be representative of low-density thermally insulating materials, e.g. glass fiber, polystyrene foams, and aerogels.[41, 42] The $C_s$ value for TIC is slightly larger than the value measured from hot disk TPS with a dry-etched sensor in Section 5.

|  |  | Ni heater (5501F / 5501 /7577) |  | Kapton layer (5501F / 5501 /7577) |
|---|---|---|---|---|
| Geometric parameters | $h_H$ (μm) | 10 | $h_{Kap}$ (μm) | 25 / 22.5 / 20 |
|  | $r_H$ (mm) | 6.403 / 6.403 / 2.001 | $r_{Kap}$ (mm) | 10 / 10 / 5.5 |
|  | $b$ (mm) | 0.213 / 0.213 / 0.25 |  |  |
|  | $m$ | 15 / 15 / 4 |  |  |
| Thermal properties | $C_{Ni}$ (MJ m$^{-3}$ K$^{-1}$) | 3.95 | $C_{Kap}$ (MJ m$^{-3}$ K$^{-1}$) | 1.55 |
|  | $k_{Ni}$ (W m$^{-1}$ K$^{-1}$) | 91.4 | $k_{Kap}^{out}$ / $k_{Kap}^{in}$ (W m$^{-1}$ K$^{-1}$) | 0.25/1.5 |
|  | $\alpha_{Ni}$ (mm$^2$ s$^{-1}$) | 23 | $\alpha_{Kap}^{out}$ / $\alpha_{Kap}^{in}$ (mm$^2$ s$^{-1}$) | 0.16/0.97 |
| Optical properties |  | Black walls | $n_{Kap}$ | 1.8 |
|  |  |  | $\beta'_{Kap}$ (10$^4$ m$^{-1}$) | 1.2 |

Table 1 (continued):

|  |  | Sample (TIA / TIB / TIC / SS)* |  | Air |
|---|---|---|---|---|
| Geometric parameters | $L_s$ (μm) | 50 | $h_{air}$ (μm) | 60 |
|  | $R_s$ (mm) | 100 | $w_{air}$ (mm) | 90 |
| Thermal properties | $C_s$ (MJ m$^{-3}$ K$^{-1}$) | 0.37 / 0.03 / 0.025 / 3.8 | $C_{air}$ (MJ m$^{-3}$ K$^{-1}$) | 0.0012 |
|  | $k_s$ (W m$^{-1}$ K$^{-1}$) | 0.023 / 0.016 / 0.033 / 13.6 | $k_{air}$ (W m$^{-1}$ K$^{-1}$) | 0.026 |
|  | $\alpha_s$ (mm$^2$ s$^{-1}$) | 0.062 / 0.53 / 1.3 / 3.6 | $\alpha_{air}$ (mm$^2$ s$^{-1}$) | 21.4 |
| Optical properties | $n_s$ | 1.07 (TIB) | $n_{air}$ | 1 |
|  | $\beta'_s$ (10$^4$ m$^{-1}$) | Varied (TIB) | $\beta'_s$ (10$^4$ m$^{-1}$) | 0 |

* For the sample, TIA = Airloy® x56, TIB = a hypothetical TI material, TIC = polystyrene foam, SS = stainless steel

## 4. Numerical Simulation Results

### 4.1 Temperature Response Sensitivity Analysis

Before the analysis of the error in the hot disk TPS measurement, we first consider the sensitivity of the average Ni heater temperature rise $\overline{\Delta T_H(t)}$ to several important parameters of the geometry and the thermal properties of the sensor and sample using COMSOL simulation, including. The sensitivity of a certain parameter $\gamma$ is defined as:

$$S_\gamma = \frac{\partial \ln(\overline{\Delta T_H(t)})}{\partial \ln(\gamma)}\bigg|_{X \neq \gamma} \approx \frac{\gamma \delta(\overline{\Delta T_H(t)})}{\Delta T \delta(\gamma)}\bigg|_{X \neq \gamma}, \qquad (7)$$

and is obtained by calculating the percentage change of the temperature increase as $\gamma$ is changed by 2% with all the other parameters fixed. The baseline parameters are shown in Table 1. The larger the $S_\gamma$ the larger change of sensor temperature when $\gamma$ is changed. Although $S_\gamma$ is not directly related to the magnitude of error due to additional step of identification with given $\overline{\Delta T_H(t)}$, $S_\gamma$ demonstrates the role of sensor in determining the heat flow. Two different types of sample: a low-$\alpha_s$ aerogel sample (TIB) and a high-$\alpha_s$ stainless steel sample are considered (see Appendix A4).

For aerogel TIB, the temperature response is highly sensitive to the $h_{Kap}$ and $C_{kap}$ in relatively short time range, indicating a considerable amount of heat diffusion and storage in the Kapton layer (especially the Kapton margin). In addition, the sensitivity to both $k_{Kap}^{in}$ and $r_{Kap}$ are high and increase with time even at long time when $S_{C_{Kap}}$ is small. This suggest that the evolution of temperature distribution near the sample/Kapton interface influenced by $k_{Kap}^{in}$ is important and the separation of the non-ideal sensor effect in the traditional identification procedure with the analytical model by removing early data points (i.e., minimizing sensitivities to Kapton properties by increasing $\tau_{cl}$) is impossible. By contrast, for stainless steel, the temperature sensitivities to $k_{Kap}^{in}$, $C_{kap}$, and $r_{Kap}$ are small, indicating a negligible amount of heat storage and in-plane heat diffusion in the Kapton layer due to the fast diffusion in the sample and short time of measurement.

4.2 Analysis of the Error in Hot Disk TPS Measurements

4.2.1 Sources of Error Due to the Finite Size of the Sensor

We propose that mainly four factors contribute to the systematic error in the experimental hot disk TPS measurement of opaque TI materials, due to the discrepancy between the ideal analytical model and the actual heat transfer process with a finite-thickness sensor. (1) The heat diffusion and storage in the Kapton insulation layer and margin at relatively short time. (2) The deviation of the temperature distribution near the sample/Kapton interface from that in the ideal heater situation at long time. (3) The poor contact between sensor and the sample which causes heating power to be trapped in the sensor and lost to the air as well as a deviation of the heating area compared with the ideal shape. (4) The heat loss to the air through the sensor side wall. These factors are never clearly discussed in prior publications. Zhang et al. attributes the error in the hot disk TPS only to the heat loss to the outer vertical side of the sensor based on numerical simulation using isothermal boundary condition.[20] Coquard et al. considered the effect of the sensor thermal mass relative to the sample without in-depth discussion of how these factors play their roles.[17]

Factor (1) and (2) occur to all samples but mainly affect TI materials which requires a long measurement time to obtain a penetration depth $d \approx 2r_H$. They are also the most important sources of error. When $\alpha_s < \alpha_{Kap}$, the fast heat diffusion in the Kapton compared with that in the sample causes a large amount of heat loss. In addition, it also causes the lateral heat diffusion in the Kapton layer and hence the effective area of heat source for the sample, i.e. the heated area in the Kapton at the sample/Kapton interface, becomes larger than the area of the ideal sensor in the analytical model where the heat source is in direct contact with the sample. Considering symmetry, this can be described by an effective sensor radius, $r_{eff}$ that correlates to the true $k_s$ which differs from the metal heat radius $r_H$ used in the identification process based on the analytical model.[††] In comparison, when $\alpha_s \gg \alpha_{Kap}$, the measurement time is too short for heat to diffuse a long distance laterally in the Kapton. This suppresses both the heat loss to the Kapton and the deviation of the $r_{eff}$ from $r_H$ and reduces the error.[‡‡] Factor (3) depends on the experimental situation and may compete with factor (2) but is difficult to simulate in numerical modeling. We only briefly consider factor (3) in the study of the interface thermal resistance in Appendix A6. Factor (4) is less important due to the low $\alpha_{air}$ and the small area of the Kapton side wall (see Appendix A5).

Both (1) and (2) lead to overprediction of the $k_{app}$ from the hot disk TPS measurement. This can be qualitatively understood with the analytical model by treating the sensor as a heat source with $r_{eff}$ varying with time and thermal mass. Consider $r_{eff}$ that changes with the change of sample/Kapton interface temperature and ignore the heat loss for now. The dimensionless temperature response $H_m(\tau, \beta)$ does not change significantly with the filling fraction of the heater area, $\beta$. With $m$ fixed, when $\beta$ increases from 1/100 to 1/$m$, i.e., from a heater with $m$ narrow rings to a full disk heater, the change of $H_m(\tau, \beta)$ is <3% at $\tau > 0.4$, see Fig. A1(d). Thus, from a given sensor temperature response curve and power input we shall have **approximately $k_{app} \propto 1/r_H$ and $k_s \propto 1/r_{eff}$** based on the model in Eq. A11. Consequently, if the $r_{eff} > r_H$, the result from the

---

[††] $r_{eff}$ should be considered as the deviation from the ideal heater averaged for all the ring heaters rather than just the outermost heater.
[‡‡] Note the cross-plane diffusion time through the Kapton layer, $h_{Kap}^2/\alpha_{Kap}^{out}$, is small since $h_{Kap} \ll r_H$.

analytical model $k_{app}$ will be larger than $k_s$. Apparently, the heat loss to the Kapton which reduces the actual power into the sample in factor (1) overpredicts of $k_{app}$.[§§]

In the next few sections, we shall use numerical simulation to quantitatively study the effect of these factors on the relative error of the identified $k_{app}$ and $C_{app}$ with respect to the input values of $k_s$ and $C_s$, i.e. $k_{app}/k_s - 1$ and $C_{app}/C_s - 1$. The contribution from sensor geometry, sensor thermal properties, and interfacial thermal resistance as well as radiative heat transfer will be analyzed.

4.2.2 Influence of the Sensor Geometry and Thermal Properties

First, we investigate the geometric parameters of the sensor. Four different sample materials including a commercial Airloy® x56 aerogel (TIA), a hypothetical TI material (TIB), a polystyrene foam (TIC), and stainless steel (SS) are studied in the investigation of the Kapton layer thickness and Kapton margin width (see Table 1). TIC and SS are both experimentally used reference materials with thermal properties well known. The $k_s$ input for TIA is from its nominal value in the material specification, and $C_s$ estimated based on the experimentally results considering the overprediction (see section 5). The four materials have the relation $\alpha_s(\text{TIA}) < \alpha_s(\text{TIB}) < \alpha_s(\text{TIC}) < \alpha_s(\text{SS})$. The in-plane diffusivity of the Kapton layer has $\alpha_s(\text{TIB}) < \alpha_{\text{Kap}}^{in} < \alpha_s(\text{TIC})$.

---

[§§] Since the identified $k_{app} \propto P_{tot}$ while the true value $k_s \propto P_{actual}$, thus $k_{app} > k_s$, if $P_{tot} > P_{actual}$ due to the heat loss.

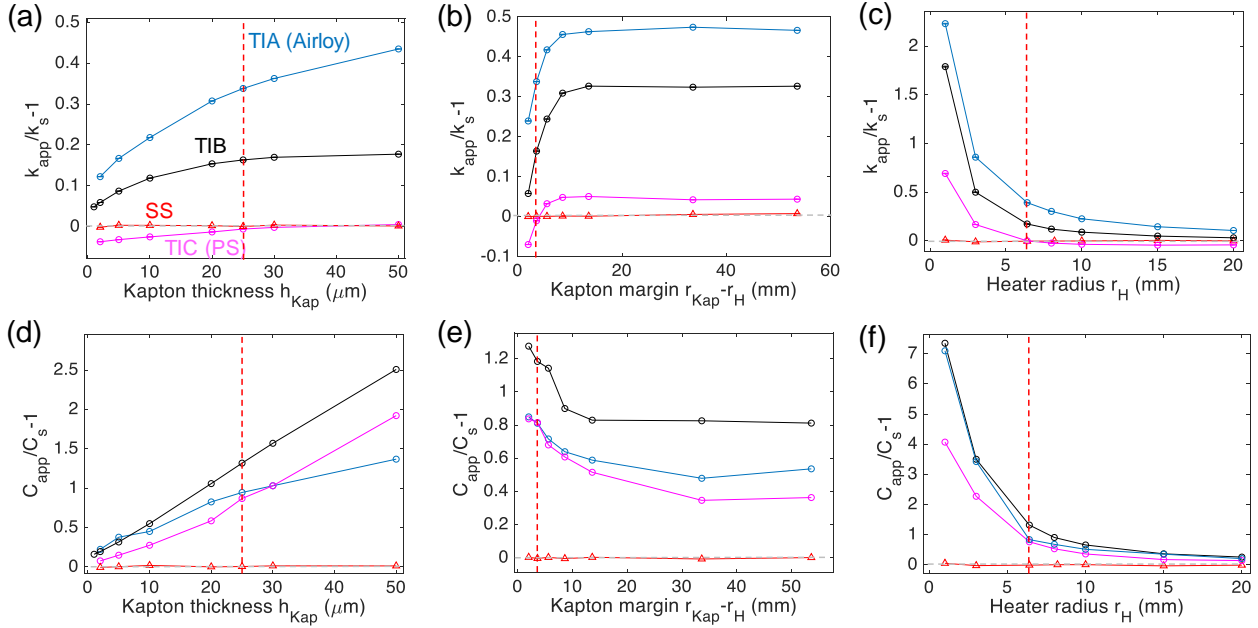

Fig. 3 Effect of the geometric factors of the sensor on the identified $k_{app}$ (top row) and heat capacity $C_{app}$ of the sample (bottom row) for the three TI materials in Table 1. Left column (a,d), the effect of the Kapton insulation layer thickness $h_{Kap}$. Middle column (b,e), the effect of the width of the Kapton layer margin by changing the radius of the Kapton insulation layer $r_{Kap}$ while holding the Ni heater radius $r_H$ fixed. Right column (c,f) the effect of the heater radius with the ring width and gap size scaled proportionally and a fixed Kapton margin width. TIA = Airloy® x56, TIB = a hypothetical TI material, TIC = polystyrene foam. Thermal diffusivities have relation $\alpha_s(TIA) < \alpha_s(TIB) < \alpha_s(TIC) < \alpha_s(SS)$. In all cases, the vertical red dashed lines indicate the baseline values of the real sensor geometry, as given in Table 1. At the same baseline values, the results of the error in $k_{app}$ and $C_{app}$ are consistent in all panels. The horizontal gray dashed lines mark the zero of the vertical axes. The trends of the errors as functions of these geometric parameters and the different behaviors for different materials are clearly seen. In general, the errors are reduced for thinner sensors with larger heater radius and smaller Kapton margin width.

Three geometric parameters are studied in Fig. 3. The baseline values of the geometric parameters are labeled by the red doted lines in the panels. In Fig. 3(a, d), the thickness of the top and bottom Kapton insulation layers, $h_{Kap}$, are changed simultaneously with the mirror plane symmetry of the system with respect to the $z = 0$ plane maintained. With decreasing Kapton thickness, the volume of the Kapton is reduced the hence the fraction of the heating power loss (storage) in the Kapton at short time is reduced. Meanwhile, a shorter length of heat diffusion along the thickness direction reduce the deviation of the temperature distribution in the Kapton layer near the sample/Kapton interface from the ideal case, bringing the $r_{eff}$ closer to the $r_H$ at long time.[***] Therefore, the hot disk TPS error is reduced with decreasing $h_{Kap}$ in general. The relative errors of the $C_{app}$ for all materials are positive numbers and in general increase as the $h_{Kap}$ increases. This is because the larger the sensor thickness, the longer the time it needs for heat to diffuse through the Kapton and reaches the penetration depth in the sample, i.e. the analytical model applies better at longer time. Thus, the linearity optimization in step (2) of the identification procedure will prefer long time range. To achieve larger $t_{max}$ with the loose constraint $0.548 < \tau_{cu} < 1$, the optimization has to reduce $\alpha_{app}$ (since $\tau_{cu} = \frac{\sqrt{(t_{max} - t_c)\alpha_s}}{r_H}$ is constrained) which leads to a larger $C_{app} = k_{app}/\alpha_{app}$, since $k_{app}$ also increases with the sensor thickness.

The relative error in $k_{app}$ in Fig. 3(a) for the 4 materials has the relation: |error(SS)| < |error(TIC)| < |error(TIB)| < |error(TIA)|. For SS, the error for both $k_{app}$ and $C_{app}$ is always nearly 0 (<0.7%) regardless of the sensor thickness in the range studied here. This is consistent with the experience that the hot disk TPS is accurate when applied to high-$k$-high-$C$ materials. In comparison, for TIA (Airloy® x56), $k_{app}$ can be ~33% higher than the real input $k_s$ when the

---

[***] This also reduce the height (and hence area) of the Kapton side wall in contact with the air and slightly reduce the heat loss to the air at long time

baseline values of the $h_{Kap}$ is used. $k_{app}$ for TIB and TIC (PS foam) with the baseline $h_{Kap}$ are 16.5% and 0.5% higher the input $k_s$ respectively. This is consistent with the trend of $\alpha_s(TIA) < \alpha_s(TIB) < \alpha_s(TIC) < \alpha_s(SS)$ and is not a coincidence. In general, as $\alpha_s$ increases, the necessary $t_{max}$ after optimization for heat to diffuse $d \sim 2r_H$ in the sample, is shorter. Thus, with increasing $\alpha_s$, the in-plane heat diffusion length in the Kapton layer is shorter, bringing $r_{eff}$ closer to $r_H$ and reducing the heat loss to the Kapton margin. Consequently, the error is smaller for sample with larger $\alpha_s$.

For the stainless steel, its high $\alpha_s(SS)$, ~2.8 times $\alpha_s(TIC)$, determines that the measurement time prove the penetration depth $d$ is 3 times shorter than that for TIC and shorter than all other materials. This leads to short optimized $t_{max}$ for SS which limits the lateral heat diffusion length and reduce the heat lost to the Kapton margin considering $\alpha_s(SS)$ is ~3.7 times $\alpha_{Kap}^{in}$. Even if $h_{Kap}$ is small, the lateral heat diffusion in the gap of Kapton does not cause the $r_{eff}$ to noticeably deviate from $r_H$. In addition, since $C_s(SS)$ is 2.5 times that of the $C_{Kap}$, and the heated volume of the sample SS is dramatically larger than the whole volume of the sensor, the heat loss to the Kapton layer is negligibly small. These reasoning also explain the negligible error in both $k_{app}$ and $C_{app}$, independent on the geometry, for SS in other panels in Fig. 3. The $C_s$ of all other TI materials are all much lower than $C_{Kap}$ (e.g. $C_s(TIC)$ is 60 times smaller than $C_{Kap}$) and hence the heat loss in the Kapton layer is still relevant even though the heated volume in the sample is larger than the that of the Kapton.

The error in $k_{app}$ for TIC when $h_{Kap}$ is small is grows negatively with decreasing $h_{Kap}$. In this case, since $\alpha_s(TIC) > \alpha_{Kap}^{in}$, the temperature rise in the Kapton near the sample/Kapton interface in the gap between Ni rings filled with Kapton becomes slower than if the gap is filled with the TIC material in the ideal sensor case. Hence $r_{eff}$ is smaller than $r_H$, leading to an underprediction of the $k_{app}$. When $h_{Kap}$ increases, the cross-plane diffusion time $h_{Kap}^2/\alpha_{Kap}^{out}$ increases allowing heat to diffuse further laterally when it reaches the sample/Kapton interface which increases $r_{eff}$. In addition, the heat loss in the Kapton layer also increase. Both effects compensate for the small $r_{eff}$ and reduce the underprediction by error canceling. Note that the error for TIC is small (max<5%) and is nearly zero when $h_{Kap}$ is large. This is partly because $\alpha_{Kap}^{in}$ is smaller but still close to $\alpha_s$(TIC) and hence the heat diffusion in the Kapton is similar to that in the sample which limits the error (see 4.2.3).

Fig. 3(b, e) show the effect of the margin width of the Kapton insulation layer $r_{Kap}$ - $r_H$ with a fixed $r_H$ (and all other geometric parameters). Due to the lateral heat diffusion in the sensor, the margin of the Kapton acts as parasitic heat loss for the measurement causing error in low-$k$ samples which require long measurement time. It also affects the size of $r_{eff}$ by changing the contribution from the outermost ring heater. The effect of the Kapton margin is not captured by previous literatures.[20] Increasing the margin width of the Kapton when it is still small increases the fraction of heat loss in the Kapton as well as $r_{eff}$ and hence worsen the overprediction of $k_{app}$. The universal trend of error saturation in Fig. 3(b) for the TI materials with a threshold at ~13 mm indicates a maximum lateral heat diffusion length determined by $d \sim 2r_H \approx 13$ mm, proving the error is indeed mainly determined by the heat loss within the lateral diffusion length in the Kapton layer margin.

Comparing the four different materials in Fig. 3(b), we still have the relation |error(SS)| < |error(TIC)| < |error(TIB)| < |error(TIA)| as in (a). This is consistent with the mechanism of lateral diffusion in the sensor and with the relative heat loss and $r_{eff}$ influenced by different measurement time scale and different materials properties discussed above. The negative error for TIC which increase in magnitude with decreasing $r_{Kap}$ indicates that the heat loss to the Kapton margin compensates the small $r_{eff}$ compared with $r_H$, supporting the conclusion for TIC in Fig. 3(a).

In Fig. 3(e), the trend of error in $C_{app}$, vs. margin width is different from that in Fig. 3(d). For low-$\alpha$ materials, when the margin width increases, the fitting shall suffer from heat loss to Kapton layer margin at long time although the optimization prefers long time range as discussed for Fig. 3(d). According to Fig. 3(e), the need to reduce the heat loss seems to dominate and the fitting prefer relatively shorter $t_{max}$ which leads to a decreasing trend in the $C_{app}$ as the Kapton margin width increases.

With a fixed margin width of the Kapton layer, we also vary the heater radius $r_H$ with the heater width $b$ and the gap size between them scaled together in Fig. 4(c, f). Increasing the radius of the whole sensor in general reduce the portion of heat loss and improve the accuracy, consistent with the conclusion by Zhang et al.[20] This is mainly because, with fixed $h_{Kap}$, the amount of heat goes to the sample scale with the heated volume times the sensor area $\sim dr_H^2 = 2r_H^3$ while the heat loss to the Kapton insulating layers scales with the volume of the Kapton $\sim r_H^2$ and the heat loss to the air at the sensor edge only scale with the area of the side wall and is proportional to $\sim r_H$. Therefore, with the margin width of the Kapton fixed, the fraction of the heat loss to the sensor relative to the heat absorbed by the sample is reduced when $r_H$ increases for all low-$\alpha$ samples. Meanwhile, since $\alpha_{Kap}^{in} > \alpha_s$(TIA, TIB) the ratio of the $r_{eff} / r_H$ slightly increases as the time of measurement time becomes longer which may worsen the overestimation. As can be seen in Fig. 4(c, f). the significant reduction of heat loss fraction dominates the competing effects (similar to the case of Fig. 4(e)), and the error in both $k_{app}$ and $C_{app}$ decrease with increasing $r_H$ for TIA and TIB. For TIC, as discussed above, since $\alpha_{Kap}^{in} < \alpha_s$(TIC), the heat diffusion near the Ni heaters is limited by the gap between rings filled with the Kapton and decrease the $r_{eff}$ relative to $r_H$ leading to underestimation of $k_{app}$. The larger the heater $r_H$, the longer time $t_{max}$ of measurement and hence the smaller the $r_{eff}/r_H$. Due to the competing effect of the reduction of heat loss fraction, the error for TIC grow negatively in the magnitude above the baseline $r_H$ then saturate.

From Fig. 3, we can see that for low-$k$ and low-$\alpha$ sample, the key to reduce the error is to reduce the thickness, the margin width of the Kapton layer and increase the Ni heater radius. These factors are in line with the intuitive understanding to reduce error by making the real sensor resemble the ideal 2D heater without any insulation layer. Of course, the radius $r_H$ should still be much smaller than the sample size $L_s$ (e.g. $L_s > 5\ r_H$) to ensure the assumption of infinite lateral sample domain in the model.

The effect of the thermal properties of the insulation layer and the heat conduction to air through the Kapton edge wall are examined in Appendix A5. We show that the hot disk TPS error is one order of magnitude more sensitive to $k_{Kap}^{in}$ than $k_{Kap}^{out}$ confirming the effect of lateral heat diffusion and $r_{eff}$. The influence of the interface thermal resistance between the sensor and sample and the thermal radiation are investigated in Appendix A6.

4.2.3 Influence of Sample Thermal Properties and the Error Correction Function

To better understand how the error in the TPS measurement depends on the sample thermal properties, we further calculate the TPS temperature with the sensor properties fixed as in Table 1 for the geometry of Kapton-5501 ($r_H = 6.403$ mm) and Kapton-7577 ($r_H = 2.001$ mm) sensors and sample $k_s$ and $C_s$ varied between 0.01 to 1.5 W m$^{-1}$ K$^{-1}$ and 0.03 to 5.6 M J m$^{-3}$ K$^{-1}$, respectively which cover the range of most low-$k$ materials. All the other parameters are fixed at the values given in Table 1. The results are then analyzed using the same identification procedure described in section 2.4. As mentioned before, since $\tau_{cu}$ is fixed in 0.548 to 1, we have $d_p \approx 2r_H$ for all sample cases and hence the condition of sample size $L_s \gg d$ is always satisfied.

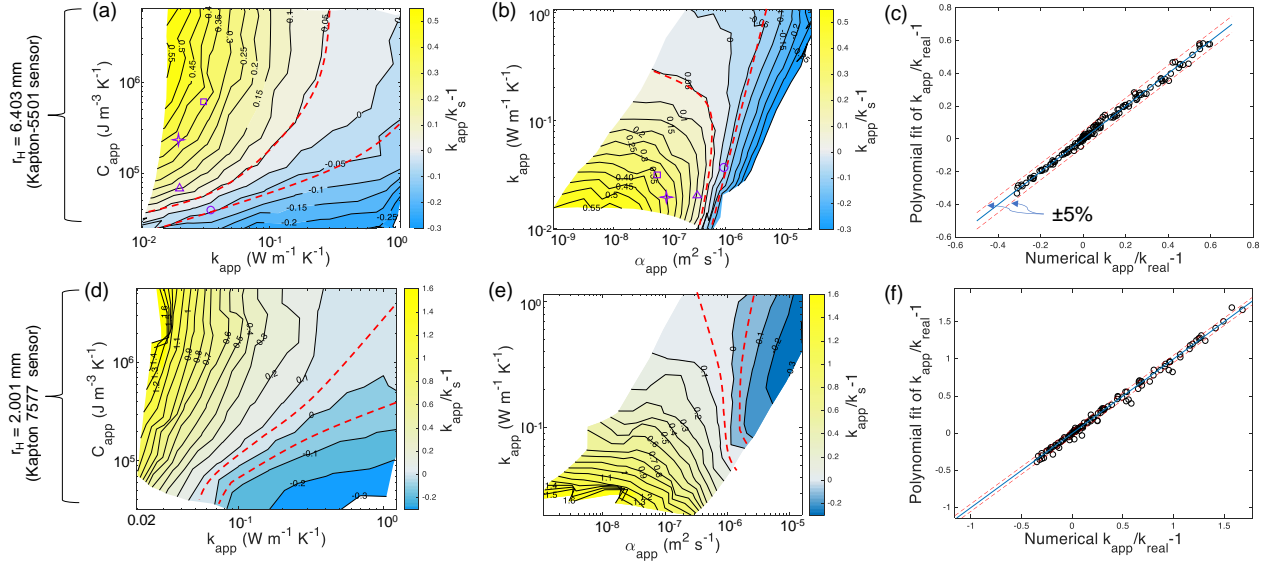

Fig. 4 Systematic error of the identified $k_{app}$ obtained with various sample thermal properties for larger-diameter (Kapton-5501, top row) and smaller-diameter (Kapton-7577, bottom row) sensors based on FEM simulation. (a, d) Contour map of the relative error vs. the $k_{app}$ and $C_{app}$. (b, e) The same data as (a) and (d) plotted vs. $k_{app}$ and $\alpha_{app}$. The points corresponding to TIA (purple open square), TIB (purple open triangle), TIC (purple open circle), and hydrophobic aerogel (purple open star) are labeled in both (a) and (b). The three regimes of the error (distinguished with boundaries of |error|≈ 5%) due to different thermal properties are marked approximately with the red dashed lines in (a) and (b). The sweet zone with nearly zero error appears as a fan-shaped regime with a vertical line at the bottom. (c, f) Comparison of the relative error identified from the numerical simulations (i.e., (a, d)) and that from the 3$^{rd}$ order polynomial fitting of Eqs. 7. The solid blue line indicates perfect agreement, while the red dashed lines indicate the situation when the polynomial fitting is ±5% away from the numerical result.

In Fig. 4, we provide a correction function that relate the apparent output from the experimental output using the pristine Kapton-5501 and Kapton-7577 sensors to the actual thermal properties of the sample. Here, we focus on the identified $k_{app}$ of the sample and plot the relative error ($k_{app}/k_s$)-1 as a function of the obtained apparent sample thermal properties $k_{app}$ and $C_{app}$. The systematic error of the experimentally studied materials of TIA, TIC, and a hydrophobic aerogel (see Table A1) are labeled in the contour map. As shown in Fig. 5(a, b, d and e), three different regimes distinguished with boundaries of |error|~5% can be identified. When the $k_{app}$ is low and $C_{app}$ is not extremely low and hence $α_{app}$ is low, the error is large positive number. This is the case for some porous TI materials (e.g. TIB, and Airloy® x56 (TIA)). In this regime the sample $α_s$ is much lower than that of the Kapton and the heat diffusion in the Kapton layer leads to large heat loss as well as the deviation of the $r_{eff}$ from the ideal Ni radius of the heater as discussed before. Both factors contribute to the overprediction of the sample $k_{app}$ with the largest error of ~60% and ~160% for sample with $k_{app}$ ~ 0.013 W m$^{-1}$ K$^{-1}$ and 0.02 W m$^{-1}$ K$^{-1}$ with Kapton-5501 and Kapton-7577 sensors respectively

On the other hand, when the sample $k_s$ and $C_s$ are both large and $α_s$ is still moderate (e.g. stainless steel) or happen to have a $α_s$ close to that of the Kapton along the in-plane direction $α_{Kap}^{in}$ = 0.97 mm$^2$ s$^{-1}$ 1 (e.g. PS foam(TIC)), the error is negligibly small, consistent with the experiment experience. The sweet zone with small error then appears as a upright fan-shaped regime in the error contour map in Fig. 7(b) with large area at the top, corresponding to high $k_s$ and moderate $α_s$ and a narrow line area at the bottom, corresponding to the vicinity of the line of $α_s \sim α_{Kap}^{in}$. This can be understood based on the arguments made for Fig. 3. For materials with high $k_s$ and $α_s$ the heat diffusion into the sample is much faster than the heat loss to the sensor and a relatively short measurement time limit the deviation of $r_{eff}$ from $r_H$. As for stainless steel, these factors suppress the error. For material with low k but $α_s \sim α_{Kap}^{in}$, like the PS foam, the error is small mainly because $r_{eff} \sim r_H$ and the relatively short measurement time reduce heat loss to the Kapton layer.

Finally, when the sample $k_s$ is large and $C_s$ is low and hence $α_s$ is large, which is rare in real materials, the contour map shows that the error can be negative number. In this regime, the time range of the analysis or measurement determined by the sample $α_s$ based on the ideal model is even shorter than that for high-$k$ high-$C$ materials. Therefore, the heat loss to the Kapton margin should not be important. However, since $α_s > α_{\text{Kap}}^{in}$ the gap and capping Kapton material limit the heat diffusion compared with the ideal case leading to $r_{\text{eff}} < r_H$ and hence an underprediction of the $k_{app}$, similar to the discussion of Fig. 3 for TIC.

To provide a convenient error correction function for engineer and materials scientist, we use a 3$^{rd}$ order polynomial function $F_c$ to fit the numerical results of the relative error $F_c = (k_{app}/k_s)-1$ in Fig. 3(a, d) with the natural logarithm of the apparent TPS results, $k_{app}$ and $C_{app}$ as variables:

$$F_c(x, y) = \sum_{i=0}^{3}\sum_{j=0}^{3} p_{ij} x^i y^j \qquad (8)$$

Here $x = \ln(k_{app})$ and $y = \ln(C_{app})$. The suitability of the polynomial fitting depicted by Fig. 3(b). and the good agreement is seen with RMSE = 0.0107 (0.0280) and adjusted $R^2$ = 0.9977 (0.9968) for Kapton-5501 (Kapton-7577) sensor. The coefficients are listed in Table 2. The coefficients of $p_{ij}$ not listed in Table 2 are zero. Note that this correction function is derived for the measurement using the Thermtest® Kapton-5501 and Kapton-7577 sensors based on the numerical and analysis process that uses the analytical model according to the standard.[9] When the correction function is applied to experimental measurement, the total sensor heat apparent in the commercial software should be set to 0 to obtain relevant $k_{app}$ to be corrected (see section 5).

Table 2 Coefficients for the polynomial fitting of the correction function for Kapton-5501 and 7577 sensors

|   | $p_{00}$ | $p_{10}$ | $p_{01}$ | $p_{20}$ | $p_{11}$ | $p_{02}$ | $p_{30}$ | $p_{21}$ | $p_{12}$ | $p_{03}$ |
|---|---|---|---|---|---|---|---|---|---|---|
| Kapton 5501 | -5.524 | 0.6417 | 1.089 | -0.01325 | -0.09402 | -0.07165 | 0.00115 | 0.00394 | 0.00358 | 0.00158 |
| Kapton 7577 | -2.523 | 0.1883 | 0.1942 | -0.2722 | -0.05616 | 0.008462 | -0.02931 | 0.01817 | 0.002816 | -0.0006668 |

## 5. Experimental Validation

### 5.1 Experimental Method

Experimental measurements are carried out with a commercial Hot Disk TPS 2200 instrument from Thermtest® to validate our theoretical predictions. To study the effect of the sensor geometry, a common Kapton-5501 (~55 µm thick) sensor and a "high temperature" Kapton-5501F (~60 µm thick) sensor (accompanied with a High temperature TPS PEEK sensor adapter) are used. Several TI materials are first measured using the two pristine sensors, then we modify the sensor by chemical etching and cutting the Kapton margin and measure the identical sample with the modified sensor after each step of the modification. (see Appendix A7 for details) We also measure the aerogel samples with steady state heat flowmeter method (HFM)[43] adapted from a guarded hot plate (LaserComp® Thermal Conductivity Instrument). A commercial HFS-4 Thin Film Heat Flux Sensor® from Omega Engineering Inc. is used.

Besides the reference materials of SRM1453 polystyrene (PS) foam and stainless steel 316 mentioned in previous sections, we measure two types of commercial strengthened aerogel: Airloy® x56 (round disk with 57 mm diameter and ~11 mm thickness) and Airloy® x103 (cuboid with 82 × 60 lateral dimensions and ~15 mm thickness).[†††] The former sample has a nominal density of 0.3 g cm$^{-3}$ while the latter has a density of 0.2 g cm$^{-3}$. Both of the aerogel samples have nominal thermal conductivity of 23 mW m$^{-1}$ K$^{-1}$ according to the manufacturer and are not transparent (presumably due to the high density or certain added opacifier). A commercial hydrophobic silica aerogel (H-aerogel) disk with ~2.6 cm diameter and 7.4 mm in thickness which is transparent and has density of 0.1 g cm$^{-3}$ is also studied.

---

[†††] It is noticed that the Airloy® x103 sample is not flat with ~1 mm difference in the height

In our TPS measurements, a moderate pressure of 1.5-2.5 psi is uniformly applied to the top surface of the bulk samples to ensure a good contact of the sensor with the sample surface. (The commercial aerogel samples have compressive yield strength of 89-94 psi.) The heating power applied is between 4 mW for the aerogels and 20 mW for the PS foam. The time of measurement, i.e. heating is 320 s for the aerogels and 20 s for the PS foam. For the stainless steel, a 0.8 W heating power is used for a 10 s measurement. These parameters are optimized such that the maximum temperature rise is between 2-5 K and the fitted maximum dimensionless time $\tau_{cu}^2$ is between 0.3 to 1 (such that the commercial software does not show any warning) and the maximum temperature residuals of the fitting is < 1 mK. The thermal penetration depth of all the measurement are checked to be smaller than any of the sample dimension.[‡‡‡] For each sample and each sensor condition, we perform at least 5 repeated measurements with the waiting time between each measurement of 10 min, sufficient for the system to cool back to room temperature.

5.2 Experimental Results

The dry etching removes most of the Kapton layer on the 5501F sensor, partially exposing the Ni heater, resulting in $h_{tot} = 20\pm3$ µm, i.e., 1/3 of the pristine sensor ($h_{tot} = 60$ µm). The Kapton margin is also removed in the RIE etching process. The wet etching reduces the sensor thickness from 55 µm to ~27 µm and is carried out in multiple repeated steps rendering a uniform thickness with variation <2 µm.[§§§] The Kapton margin of the wet-etched sensor is cut by from ~3.6 mm to <1 mm. Measurement is performed after each step of the wet etching on the same sample.

---

[‡‡‡] $t_{max}$ in the identification process for the H-aerogel is chosen to be smaller than 1 to avoid the sample boundaries.

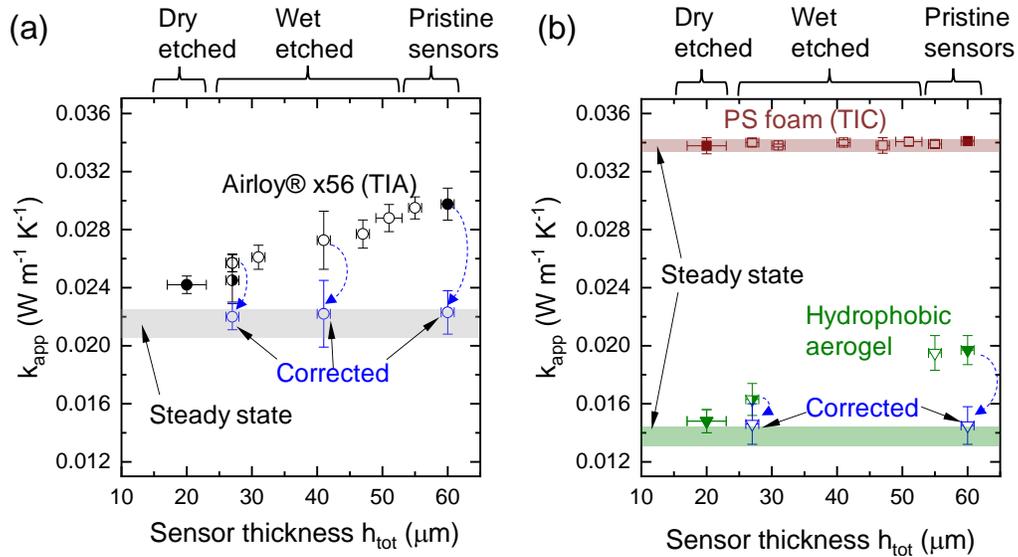

Fig. 5 Experimental validation of the effect of modifying the sensor insulation layer and correction factors in hot disk TPS measurement of TI materials. (a) Results for Airloy® x56 by hot disk TPS (black circles) and steady state heat flow meter (grey band). (b) Results for hydrophobic silica aerogel (H-aerogel) by hot disk TPS (green triangles) and steady state heat flow meter (green band). $k_{app}$ for these TI materials decreases and approaches the steady state results as the sensor thickness reduces. The hot disk and steady state results for the PS foam agree well and are shown as dark red squares and band respectively as shown in (b). The steady state method results shown as band have width spanning mean ± one standard deviation. In both (a, b), open symbols represent data obtained using one Kapton-5501 sensor ($h_{tot}$ = 55 µm, $r_H$ = 6.403 mm)repeated through various stages of multi-step wet etching, and solid symbols are results from one Kapton-5501F sensor ($h_{tot}$ = 60 µm, $r_H$ = 6.403 mm) before and after dry etching. The blue open symbols corresponding to experimental data for the Airloy® x56 and H-aerogel samples measured with different sensor thickness show the results of correcting the commercial software-determined values using the

---

[§§§] Wet etching can hardly further decrease the thickness since there is a thin layer of adhesive or polymer coating on the Ni heater which is almost intact in the Kapton etchant (although it is removable by RIE).

polynomial correction function (based on Eq. (8) for the pristine sensor, and a separate calculation for the wet etched sensor with other reduced thickness, see main text). The discussion about the sensor heat capacity, $C_{\text{sensor}}$, is in the main text and the Appendix A8. The two half-filled symbols are the result from measurements using the wet etched sensor with the Kapton margin trimmed from ~3.6 mm to <1 mm in width. Error bars for uncorrected data indicate standard deviation of repeated TPS measurements while the error bars in $k_{\text{app}}$ for the corrected data include the error in the polynomial fitting of the correction function.

The sample $k_{\text{app}}$ measured using the two pristine Kapton insulated sensors (Kapton-5501 and Kapton-5501F) are summarized in Table A1. Note that $C_{\text{app}}$ of the aerogel samples are one order of magnitude higher than the PS foam MJ m$^{-3}$ K$^{-1}$. The uncertainty range in these apparent results for each sample are standard deviations of ~6-10 repeated measurements. Since the two Airloy® aerogels have high density and are not as transparent as the classical silica aerogel (i.e. absorption > $1.3 \times 10^4$ m$^{-1}$), the error due to radiation should be small. For the transparent hydrophobic aerogel, the radiative heat transfer occurs in both the hot disk TPS and HFM measurement and hence both methods should give similar total effective thermal conductivity.

In Fig. 5, we plot the hot disk TPS measurement results with commercial software-determined values using modified sensors for three representative materials: Airloy® x56 (black circles), SRM 1453 PS foam (dark red squares), and the hydrophobic silica aerogel (H-aerogel, green triangles). The steady state HFM measurement results are also included for comparison, shown as the horizontal bands with corresponding color which include uncertainty. The measured $k$ for the SRM 1453 PS foam from HFM is consistent with the reference data from NIST (0.033 m$^{-1}$ K$^{-1}$),[44] and the $k_{\text{app}}$ for Airloy® x56 from HFM is consistent with the value provided by the manufacture (0.023 m$^{-1}$ K$^{-1}$). Although there is no reference data for the H-aerogel, the $k$ from HFM is within the range for such silica aerogels in the literature.[45] The hot disk TPS data for Airloy® x103 (Table A1) is found to be affected by its non-flat surface and is not shown in Fig. 5.

The hot disk TPS data of $k_{app}$ determined by the commercial software decreases as the sensor thickness reduces and approaches the HFM results in Airloy® x56 and the H-aerogel sample. The $k_{app}$ for Airloy® x56 is reduced from 0.0296 W m$^{-1}$ K$^{-1}$ to 0.0242 W m$^{-1}$ K$^{-1}$ and $k_{app}$ of the H-aerogel is reduced from 0.0196 W m$^{-1}$ K$^{-1}$ to 0.0148 W m$^{-1}$ K$^{-1}$ when the Kapton layer thickness is reduced from 60 to ~20 µm through dry etching. By contrast, the measured $k_{app}$ values are unaffected by sensor thickness for the PS foam (i.e. TIC) sample. The trend of $k_{app}$ vs. $h_{tot}$ for the aerogels is consistent with our numerical simulation prediction (Fig 3(a)). The consistency between the hot disk TPS and steady state HFM results for the PS foam is confirms our numerical calculation which predict an error <1% for $h_{tot}$ = 20 - 60 µm ($h_{Kap}$ =10-25 µm) for the PS foam. The measured $k_{app}$ values of all aerogel samples (including Airloy® x103, see Table A1) are clearly reduced by more than their uncertainties and the typical maximum error of 5%,[9] and approach the results from HFM measurements.

The measured $k_{app}$ of the stainless steel sample (see Table A1) using the wet etched sensor with 27 µm thickness is 13.60±0.07 W m$^{-1}$ K$^{-1}$, virtually indistinguishable from the measurement using the pristine sensor, 13.56±0.06 W m$^{-1}$ K$^{-1}$.[****] The results from the dry etched sensor for each material follow the same trend of the thickness dependent $k_{app}$ for the wet etched sensor. Further decreasing the Kapton layer margin width of the wet etched sensor by manually trimming it also improves the accuracy, as can be seen in the Airloy® x56 sample (half-filled circle in Fig.

---

[****] The stainless steel sample could not be measured with the dry etched sensor since the Ni heater is partially exposed in the dry-etched sensor.

5(a)). Although it is hard to quantify the amount of the remaining Kapton margin, the trend is consistent with our numerical prediction in Fig 3(b).

Moreover, we present the corrected thermal conductivity of the Airloy® x56 and the H-aerogel sample based on the result of the pristine sensor and etched sensor using the corresponding correction function. Eq. 8 is used for results from the pristine 60 µm sensor. The correction process is applied to the experimental $k_{app}$ of Airloy® x56 measured with wet etched sensor with reduced thickness to further validate the method used to obtain Eq. 8.†††† The corrected results show very good agreement with the HFM measurement. Specifically, whereas the raw values measured with the pristine sensor are in error by 35% and 40% larger than the HFM reference values for the Airloy® x56 and the H-aerogel, respectively, the correction factor greatly reduces those errors to <2% and <4% (well within the mutual uncertainty of TPS and HFM). Such results quantitatively support the validity of our numerical model and the corresponding correction functions and verify our prediction that the reduction of the Kapton thickness and margin width can effectively reduce the systematic error especially for low-$k$ materials. It is also the first time in research the accuracy of the TPS is improved via experimental method.

Due to the finite thermal mass of sensor which reduces the total power from the heater to the sample, the commercial hot disk software includes the total sensor heat capacity $C_{sensor}$ in the unit of mJ K$^{-1}$ as an input parameter for fitting. We note that tuning $C_{sensor}$ in the commercial software cannot reduce the systematic error in the aerogel samples: even using $C_{sensor}$ = 0 mJ K$^{-1}$ for the pristine sensor can only reduce $k_{app}$ by 1%, giving error >30%. See discussed in Appendix A8.

---

†††† Parametric sweeps of sample thermal properties similar to Fig. 4 is conducted in COMSOL at the vicinity of the values of the Airloy® x56, then 3$^{rd}$ order polynomial fittings of the relative errors as in Eq. 8 are conducted which are then used to correct the experimental $k_{app}$ values. Since these correction functions are not universal for pristine sensors, they are not presented here.

6. Conclusion

As a widely used method for thermal properties characterization, the hot disk transient plane source (TPS) technique is of great importance for materials evaluation and development. Despite its convenience and versatility, for low thermal conductivity (*k*) bulk materials, the low accuracy of the TPS method is known to many researchers. However, the lack of a systematic investigation of the reason behind such issue as well as the lack of a practical engineering method to improve it has limited the reliability and applicability of the hot disk TPS method for such materials. In this work, the influence of the hot disk sensor geometry and thermal properties on the measurement error are carefully studied for a series of different sample materials. We reveal that the error is correlated with the lateral heat diffusion and loss in the insulation layer as well as the deviation of temperature distribution in the insulation layer from the ideal case near its interface with the sample. Correction functions for common Kapton-5501 and 7577 sensors are provided to improve the accuracy of practical hot disk TPS measurement of low-*k* materials. With experimental modification of the sensor, we show that the reduction the thickness and the insulation layer margin width of the sensor (Kapton-5501) is an effective way to reduce the error due to both mechanisms. Such results are important to further improve the accuracy and reliability of the TPS method for low-*k* material and can be employed to design new type of sensor with a broader application range.

Acknowledgement

This work was supported by Energy Efficiency and Renewable Energy, Building Technologies Program, of the U.S. Department of Energy under Contract No. DEAC02-05CH11231. The authors would like to thank Dr. Howdy Goudey for the help on heat flowmeter measurements. Q. Zheng would like to thank Dr. Jie Zhou for the help in numerical simulation.

Appendix

A1. Improved Analytical Model

In the following derivation, we only consider $\tau = \frac{\sqrt{t\alpha_s}}{r_H}$ instead of $\tau_c$ omitting the time correction $t_c$ in the theoretical model. To be consistent with literature and for simplicity, $D_m(\tau_c)$ is written as $D(\tau_c)$ from here below. $\tau$ shall be replaced with $\tau_c$ in the equations if instrument delay is to be considered in the more practical situation. In Eq. 4, when the variable $\sigma \to 0$, we have $\frac{lk}{2m^2\sigma^2} \to \infty$. Thus, the modified Bessel function $I_0$ can be asymptotically expanded as

$$I_0(z) \sim \frac{e^z}{\sqrt{2\pi z}}\{1+\frac{1}{8z}+\frac{9}{128z^2}+O(z^{-3})\}$$ with large $|z|$ and arg(z) = 0 (since $\frac{lk}{2m^2\sigma^2}$ is a real number).[36]

Substitute this expression in (4), for small $\tau_c \to 0$, we have

$$D(\tau) \sim \int_0^\tau \sigma^{-2}\left\{\sum_{l=1}^m l\sum_{k=1}^m k\exp\left(-\frac{(k-l)^2}{4m^2\sigma^2}\right)\sqrt{\frac{m^2\sigma^2}{\pi lk}}\left[1-\frac{m^2\sigma^2}{4lk}+\frac{9}{32}\left(\frac{m^2\sigma^2}{lk}\right)^2+O(\sigma^6)\right]\right\}d\sigma, \quad (A1)$$

For $l = k$,

$$D(\tau) \sim \int_0^{\tau_c} \sum_{l=1}^{m} \sum_{l=1}^{m} \frac{ml}{\sqrt{\pi}} \frac{1}{\sigma} \left[ 1 - \frac{m^2\sigma^2}{4lk} + \frac{9}{32}\left(\frac{m^2\sigma^2}{lk}\right)^2 + O(\sigma^6) \right] d\sigma$$

$$\sim \int_0^{\tau_c} \sum_{l=1}^{m} \sum_{l=1}^{m} \frac{ml}{\sqrt{\pi}} \left[ \frac{1}{\sigma} - \frac{m^2\sigma}{4lk} + O(\sigma^3) \right] d\sigma$$

(A2)

The first term in the square bracket becomes the integral of $1/\sigma$, which diverges as $\ln(\sigma)$ as $\sigma \to 0$ when it is evaluated at the lower limit.

To solve the diverging problem, we refer to the original Green's function method considering *m* equally spaced concentric ring heater with finite ring width *b* instead of infinitely narrow rings. and radius of $r_j = (j/m)r_H$ $j = 1, 2, \ldots m$ with $b \leq r_H/(m)$. Note that when $b = r_H/(m)$, the heater is a full disk.

We first write the Greens' function solution of temperature in a cylindrical coordinate for heat conduction with a spatially uniform step function heat source of a single ring heater. The heater has an outer radius of $r_j$ and width of *c* and the heating power has a distribution *P(z, r, θ, t)* $= P_0 \delta(z) u(r_j - |\vec{r}|) u(|\vec{r}| - (r_j - c)) u(t - t')$, where $P_0$ is the areal power density (with unit W m$^{-2}$), *u(r)* is the Heaviside step function and *δ(z)* is the Dirac delta function.

$$\Delta T_s(z, r, \theta, t) = \iiint_V \int_0^\infty P(z-z', r-r', \theta-\theta', t-t') G(z', r', t') r' dr' dz' d\theta' dt'$$

$$= \iiint_V \int_0^\infty \frac{P_0}{\rho_s c_s} \frac{\delta(z-z') u(r_j - |\vec{r} - \vec{r}'|) u(|\vec{r} - \vec{r}'| - (r_j - c)) u(t-t')}{(4\alpha_s \pi t')^{3/2}} \exp\left(-\frac{(z'^2 + r'^2)}{4\alpha_s t'}\right) r' dr' dz' d\theta' dt'$$

$$= \iint_{S_j} \int_0^t \frac{P_0}{\rho_s c_s} \frac{1}{(4\alpha_s \pi t')^{3/2}} \exp\left(-\frac{z^2 + r'^2}{4\alpha t'}\right) r' dr' d\theta' dt'$$

(A3)

where $\rho_s$ is the sample density, $c_s$ is the sample specific heat, $V$ is the whole 3D space and $S_j$ is the 2D area determined by the heat source dimensions $r_j$, $\theta$, and $r$. Since only the temperature change in the area covered by the ring heater matters, we set $z = 0$. Let $v = \left(\dfrac{r'^2}{4\alpha_s(t')}\right)$, then

$$\Delta T(z=0,r,\theta,t) = \iint_{S_j} \int_{t_0}^{\infty} \dfrac{P_0}{\rho_s c_s} \left(\dfrac{v}{\pi(r'^2)}\right)^{3/2} \dfrac{r'^2}{(4\alpha_s v^2)} \exp(-v) r' dv dr' d\theta' \tag{A4}$$

where $t_0 = \left(\dfrac{z^2 + r'^2}{4\alpha_s(t)}\right)$. Hence, we have:

$$\Delta T(z=0,r,\theta,t) = \dfrac{P_0}{4\pi k_s} \iint_{S_j} \left(\dfrac{1}{\pi(r'^2)}\right)^{1/2} \Gamma(1/2, \dfrac{r'^2}{4\alpha_s t}) r' dr' d\theta' = \dfrac{P_0}{4(\pi)^{3/2} k_s} \iint_{S_j} \Gamma(1/2, \dfrac{r'^2}{4\alpha_s t}) dr' d\theta',$$

(A5)

where $\Gamma(1/2, x) = \int_{x}^{\infty} v^{1/2-1} dv \exp(-v) dv$ is the incomplete Gamma function which equals $\sqrt{\pi}/2$ at $x = 0$ and monotonically approaches 0 as $x$ increases with no singular point in $[0, \infty)$.

Let $q = r/r_H$, $p = r'/r_H$, $\beta = b/r_H$ and $\tau = \dfrac{\sqrt{t\alpha_s}}{r_H}$. Apparently, when $\beta = 1/m$, the heater is a full disk. When $\beta \to 0$, which corresponds to the impractical infinitely narrow ring heaters situation, the solution will be the same as the original model and shall diverge. By considering all situation of the coordinate of the point $(r, \theta)$ relative to the finite-width ring heater which influence the area of integral, $S_j$, we can derive a piecewise function for $\Delta T_s$ in the space:

For $q < j/m - \beta$

$$\Delta T_s(\frac{j}{m},\beta,q,\theta,\tau) = \frac{r_H P_0}{2k_s(\pi)^{3/2}} \int_0^{\pi} \int_{F_+(j/m-\beta,q,\theta')}^{F_+(j/m,q,\theta')} \Gamma(1/2, \frac{p^2}{4\tau^2}) dp d\theta',$$

For $j/m - \beta \leq q \leq j/m$

$$\Delta T_s(\frac{j}{m},\beta,q,\theta,\tau) = \frac{r_H P_0}{2k_s(\pi)^{3/2}} \left\{ \begin{array}{l} \int_0^{\arcsin((\frac{j}{m}-\beta)/q)} [\int_{F_+(\frac{k}{m}-\beta,q,\theta')}^{F_+(\frac{j}{m},q,\theta')} \Gamma(1/2, \frac{p^2}{4\tau^2}) dp + \int_0^{F_-(\frac{j}{m}-\beta,q,\theta')} \Gamma(1/2, \frac{p^2}{4\tau^2}) dp] d\theta' \\ + \int_{\arcsin((\frac{j}{m}-\beta)/q)}^{\pi} \int_0^{F_+(\frac{j}{m},q,\theta')} \Gamma(1/2, \frac{p^2}{4\tau^2}) dp d\theta' \end{array} \right\},$$

For $j/m < q$

$$\Delta T_s(\frac{j}{m},\beta,q,\theta,\tau) = \frac{r_H P_0}{2k_s(\pi)^{3/2}} \left\{ \begin{array}{l} \int_0^{\arcsin((\frac{j}{m}-\beta)/q)} [\int_{F_+(\frac{k}{m}-\beta,q,\theta')}^{F_+(\frac{j}{m},q,\theta')} \Gamma(1/2, \frac{p^2}{4\tau^2}) dp + \int_{F_-(\frac{k}{m},h,\theta')}^{F_-(\frac{j}{m}-\beta,h,\theta')} \Gamma(1/2, \frac{p^2}{4\tau^2}) dp] d\theta' \\ + \int_{\arcsin((\frac{j}{m}-\beta)/q)}^{\arcsin(\frac{j}{qm})} \int_{F_-(\frac{j}{m},q,\theta')}^{F_+(\frac{j}{m},q,\theta')} \Gamma(1/2, \frac{p^2}{4\tau^2}) dp d\theta' \end{array} \right\}$$

(A6)

where $F_{\pm}(x,q,\theta') = q\cos\theta' \pm \sqrt{x^2 - \sin^2\theta' q^2}$. Since the limits of the integrals are all finite, and the incomplete Gamma functions in the integrand are bounded for all real positive variables, the temperature rise does not diverge at any point in space and time. Obviously, the temperature integrated form of $\Delta T_s(\frac{j}{m},\beta,q,\theta,\tau)$ is independent on the angle $\theta$, which is as expected. Define

$$\Delta T_s(\frac{j}{m},\beta,q,\theta,\tau) = \frac{r_H P_0}{2k_s(\pi)^{3/2}} \overline{\Delta T_s}(\frac{j}{m},\beta,q,\theta,\tau) \tag{A7}$$

where $\overline{\Delta T_s}(\frac{j}{m},\beta,q,\theta,\tau)$ is a dimensionless piecewise function containing terms in the bracket in Eq. A6. The temperature rise at position $q$ is the sum of contribution from all ring heaters, i.e.,

$$\Delta T_s(\beta,q,\theta,\tau) = \sum_{j=1}^{m}\Delta T_s(\frac{j}{m},\beta,q,\theta,\tau) = \frac{r_H P_0}{2k_s(\pi)^{3/2}}\sum_{j=1}^{m}\Delta T_s(\frac{j}{m},\beta,q,\theta,\tau). \quad (A8)$$

By further taking the average of the temperature inside each ring with contribution of all rings, we can obtain the average temperature rise of the whole heater containing $m$ rings as:

$$\overline{\Delta T_s(\tau)} = \frac{r_H^2}{\pi r_H^2 (1+m(1-\beta))\beta}\sum_{l=1}^{m}\int_{0}^{2\pi}\int_{\frac{l}{m}-\beta}^{\frac{l}{m}}\Delta T_s(\beta,q,\theta,\tau)qdqd\theta$$

$$= \frac{P_{tot}}{2k_s r_H(\pi)^{5/2}[(1+m(1-\beta))\beta]^2}\sum_{l=1}^{m}\sum_{j=1}^{m}\int_{\frac{l}{m}-\beta}^{\frac{l}{m}}\Delta T_s(\frac{j}{m},\beta,q,\theta,\tau)qdq \quad (A9)$$

Define a dimensionless function

$$H_m(\tau,\beta) = \frac{1}{2\pi[(1+m(1-\beta))\beta]^2}\sum_{l=1}^{m}\sum_{j=1}^{m}\int_{\frac{l}{m}-\beta}^{\frac{l}{m}}\Delta T_s(\frac{j}{m},\beta,q,\theta,\tau)qdq. \quad (A10)$$

Then the average temperature rise can be written as

$$\overline{\Delta T_s(\tau)} = \frac{P_{tot}}{(\pi)^{3/2}k_s r_H}H_m(\tau,\beta) \quad (A11)$$

If the sample temperature response is accurately known, the derivation of $H_m(\tau,\beta)$ and $D_m(\tau)$ based on Eq. A11 and Eq. 3 should give the same magnitude. However, Substituting in $\beta = 1/(2m)$ for the real sensor and $\beta = 1/100$, we plot the dimensionless temperature rise $H_m(\tau,\beta)$ for $m = 10$ and 15 in Fig. 1 in comparison with the $D_m(\tau)$ in Eq. 4 calculated with different lower bound (lb) of the integration and $D_m(\tau)$ digitized from literature. In addition, we compare these results from the analytical model to a finite element numerical simulation of the same problem as in the derivation of Eq. A8 (i.e. finite width 2D concentric ring heater at $z = 0$ plane in a large enough 3D space filled with the sample material) using COMSOL Multiphysics®. Apparently, since the integral of Eq. 4 diverges as $\tau$ approaches 0, the results are strongly dependent on the lower limit of the integration. In addition, the results of the $D_m(\tau)$ from references show inconsistency with each other and deviate from the numerical simulation, presumably due to the discrepancy in the choice of the cutoff of $\tau$ and the approximation used in the integral. By contrast, the result of Eq. (A10) shows good agreement with the numerical simulation within 0.5%. In Fig. A1 (d), we also show $H_m(\tau,\beta)$ with $\beta = 1/m$ to demonstrate the dimensionless temperature response of a full disk heater. It is also confirmed that different choice of $m$ gives consistent result for the full disk heater.

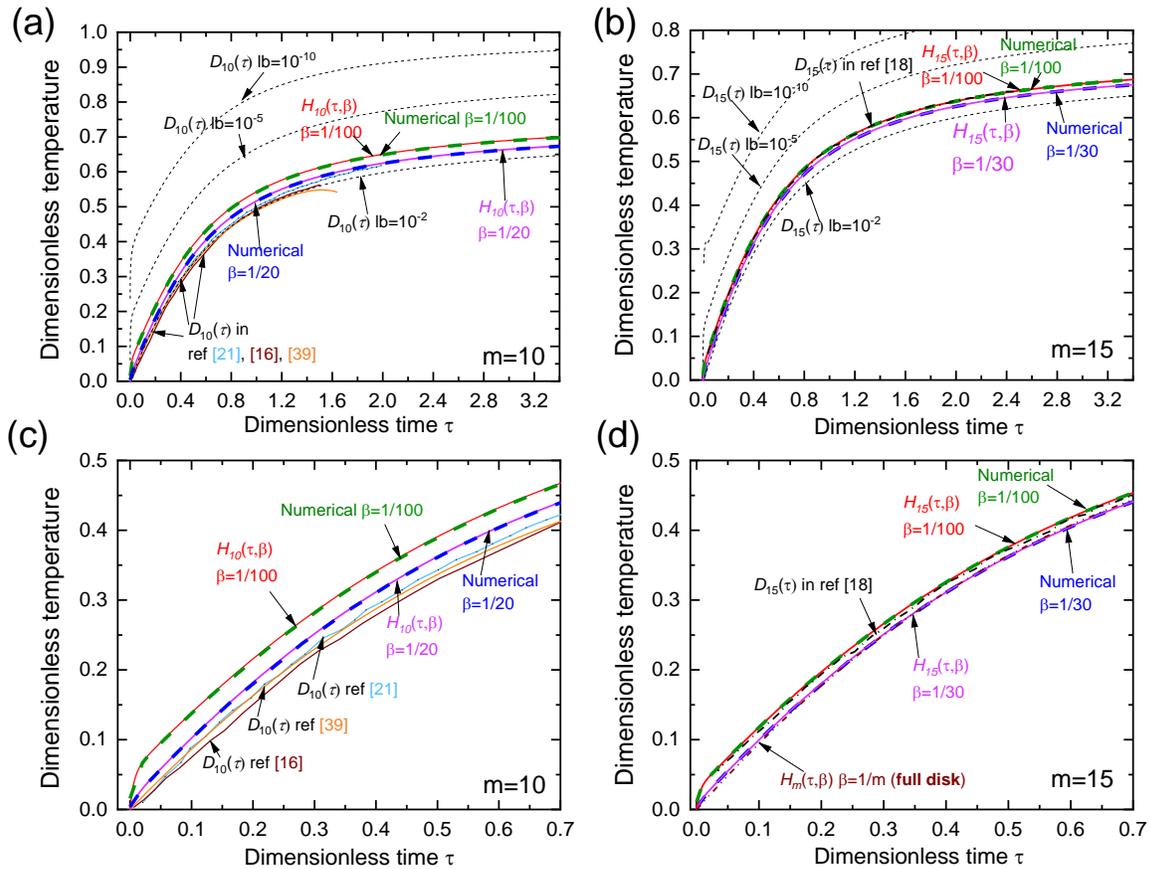

Fig. A1. (a, b) Comparison of the new analytical model for the sensor temperature response ($H_m$ from Eq. A10: solid red and magenta lines, for different values of $\beta$) with the traditional model ($D_m$ from Eq. 4: black dotted lines for three different lower bounds (lb) of the integral). Both models depend on the number of rings, $m$, and the dimensionless time, $\tau$, while the new model also takes into account the effect of finite heater width $b$, as nondimensionalized through $\beta = b/r_H$. Two different values of $\beta$ are considered in each case of $m$. Dimensionless temperature from numerical calculation is also shown (green and blue dashed lines) which agree well with the new model. The original model implicitly assumes $\beta \to 0$. Due to a known artifact in the original model as discussed in the text, it exhibits a strong but nonphysical vertical shift depending on the lower bound (lb) used in the $\sigma$ integral in Eq. 4, which diverges as $\sigma_{lb} \to 0$. Solid colored lines depict several other literature results for Eq. 4 which dealt with the divergence issue in various ways, see main text. (c, d) Same as in (a, b) with magnified scale (the black dashed lines of the calculated $D_m$ are removed for clarity). The result of the new model with $m = 15$ and $\beta = 1/30$ is used in the analysis of the TPS data from numerical calculation in other sections in the main text. The dimensionless temperature with $\beta = 1/m$ which corresponds to a full disk heater is also shown in (d) (brown dashed line).

We do not further develop analytical model that considers the finite thickness and thermal mass of the sensor since the goal of this work is to study the error in the analysis of hot disk TPS measurement using commercial software[38] which uses $D_m(\tau_c)$ in Eq. 4 (corrected in certain way so that matches $H_m(\tau_c, \beta)$) and to provide relevant correction factor. Hence, it is not useful for us to build more sophisticated analytical model here. In the later hot disk TPS data analysis for different sensors types, the new analytical result of $H_m(\tau_c, \beta)$ with corresponding $m$ and $\beta = 1/(2m)$ which is consistent with $D_m(\tau_c)$ in Ref [11] when $m = 15$ is used, replacing $D_m(\tau_c)$ in Eq. (6).

A2. Identification Procedure in This Work

As mentioned in the main text, a home-developed program is used to identify thermal properties from the hot disk TPS data generated from FEM simulation. This program is validated by analyzing raw hot disk experimental data as well as FEM simulation. For low-$k$ materials the time correction $t_c$ defined as hot disk hardware delay are small and can often be ignored when the measurement time is long according to the Hot Disk TPS® manual and the standard.[9, 38] For the analysis of the experimental data, we set the initial value of $t_c = 0$ and find that the fitted $t_c$ is less than 0.1% of $t_{max}$ after fitting. For the FEM data analysis, $t_c$ is fixed at 0 in the identification procedure for the numerical simulation data.

The initial value of the $\tau_{cl}$ for both the experimental and FEM data analysis is set to a relatively small number (~0.15 – 0.2) such that the fitting procedure slightly increases $\tau_{cl}$ after the final optimization around which the $R^2$ is maximized and the fitting $k_{app}$ is stable. On the other hand, the achievable magnitude of $\tau_{cu}$ can easily exceed 1 since $t$ in the simulation is typically set to be long enough. Therefore, only lower limit of the dimensionless time $\tau_{cl}$ is adjusted in the fitting (Fig. 2(b)). $\tau_{cu}$ changes in accordance with $\alpha_{app}$ and is chosen to be the minimum of 1 and the $\tau_c$ value that correspond to the last data point in time. A constraint of $0.548 < \tau_{cu} < 1$ according to the standard[9] is posed in the optimization process mainly to ensure the value of $\alpha_{app}$ will not lead to $\tau_{cu} < 0.548$. This home-developed program is expected to give similar or identical result of the $k_{app}$ and $C_{app}$ from the commercial software for a given set of data.

We also find that the initial value of $\tau_{cl}$ has negligible influence of the fitting result for $k_{app}$ and $\alpha_{app}$ as long as it is smaller than 0.5 and the optimization typically gives a fitted value of $\tau_{cl}$ around 0.2 which is similar to the data range after user adjustment used in the commercial software. To study if the optimization of $R^2$ is enough, we intentionally add normally distributed noise to the numerical data in the test run to better mimic experimental results. It is found that when the noise is not too large ($\leq 10^{-3}$ of the maximum temperature rise, which is common for experimental data), an extra step of optimization of the normality of the residual using a Shapiro–Wilk test by adjusting $\tau_{cl}$ and $\alpha_{app}$ after the 2nd step of optimizing the $R^2$ causes less than 2-3% difference in the final result of $k_{app}$ and $\alpha_{app}$. Thus, we do not include this extra step in the identification process.

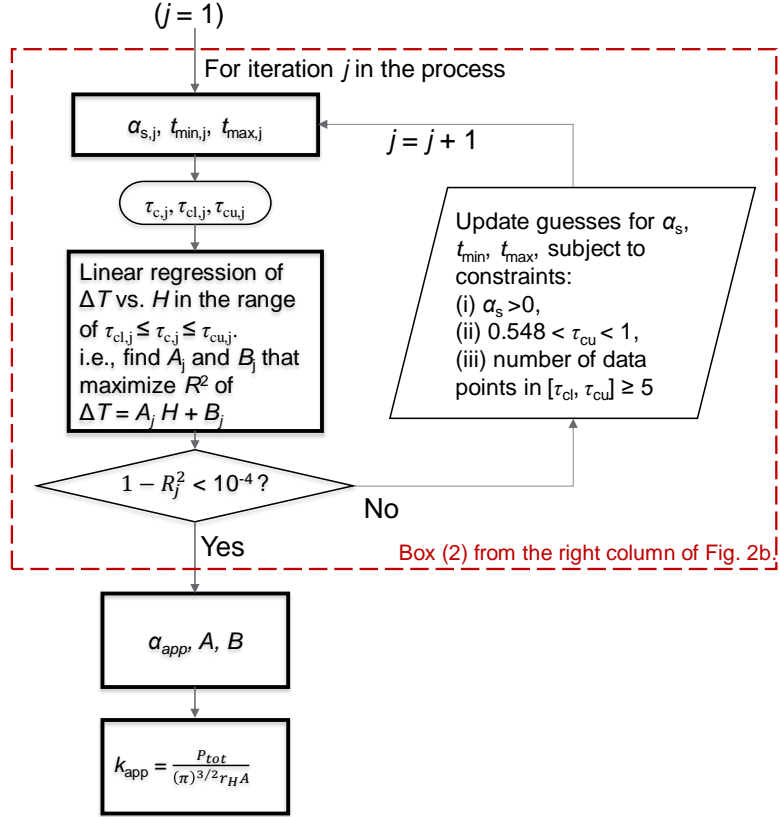

Fig. A2. Additional details about the optimization process to identify the apparent thermal properties from FEM generated data of this work in Fig. 2b (right column). $t_c$ is fixed at 0.

To further validate the home-developed code, we analyze the experimental data of standard reference materials: SRM1453 polystyrene (PS) foam and stainless steel, using this code and compare the result with the calculation output of the commercial Hot Disk Thermal Constants Analyzer®. The agreement between the $k_{app}$ derived from the two methods is excellent. For the PS foam and stainless steel, the Hot Disk software gives 0.0339±0.0003 W m$^{-1}$ K$^{-1}$, and 13.56±0.08 W m$^{-1}$ K$^{-1}$ respectively, while the code home-developed gives 0.0336±0.0003 W m$^{-1}$ K$^{-1}$ and 13.60±0.08 W m$^{-1}$ K$^{-1}$, respectively where the uncertainties are from the standard deviations. The $α_{app}$ and $C_{app}$ of the two materials obtained from the two methods are also consistent within 3%.

As a demonstration of the identification process, we show the fitting and analysis of the numerical data for an ideal hot disk sensor with zero thickness (i.e. concentric 2D rings with finite width) in Fig. A3 using the home-developed code for sample with varied input values of $k_s$ from 0.01 to 1 W m$^{-1}$ K$^{-1}$ and a fixed sample heat capacity $C_s$ of 0.03 MJ m$^{-3}$ K$^{-1}$. The fitting results match with the numerical data in the entire time range. The root means square error (RMSE) and the coefficient of determination $R^2$ after optimization are around 10$^{-3}$ K and >0.99992 respectively. The maximum error of the identified $k_{app}$ and $α_{app}$ is ~0.4% and ~2% respectively which confirms the reliability of the identification procedure we developed here. As expected for an ideal sensor, the fitted $τ_{cl}$ is ~10$^{-4}$, close to 0 (not shown).

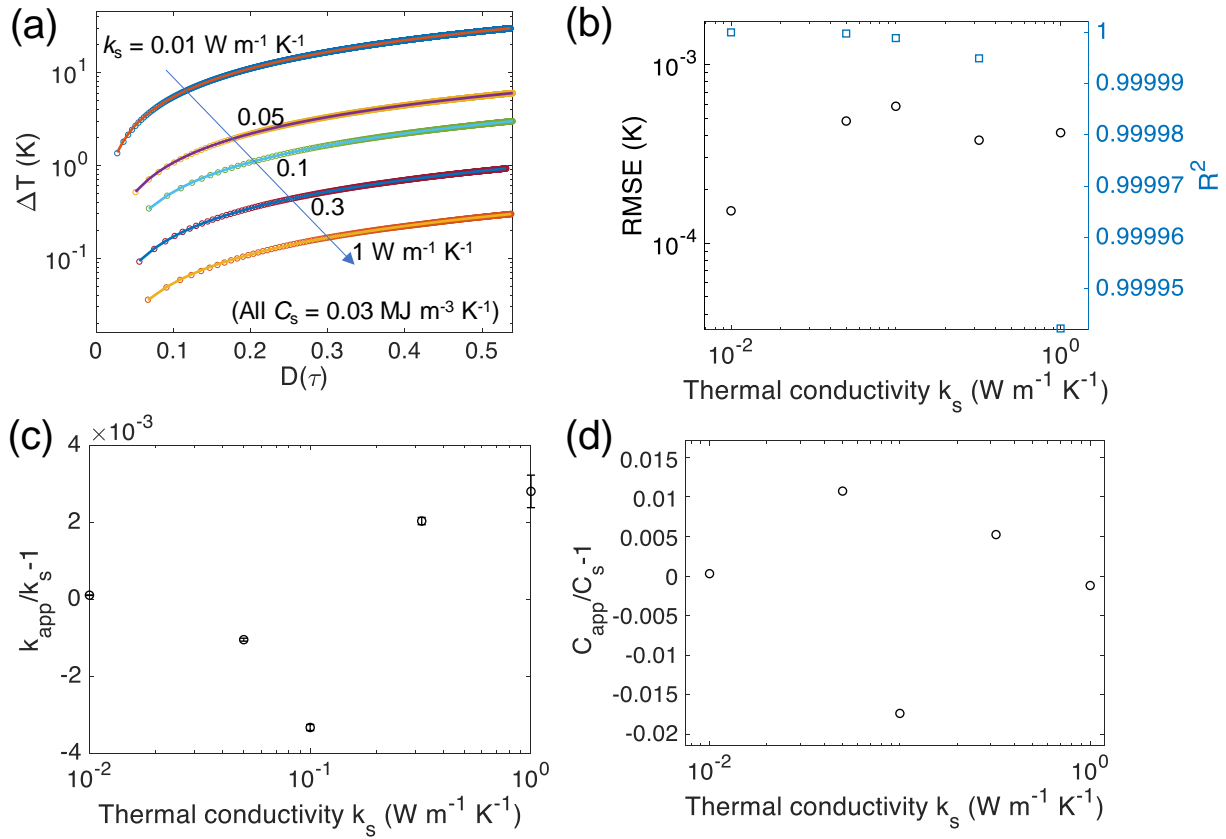

Fig. A3 Numerical simulated TPS data and the results from the 2-step identification procedure used in this work of an ideal sensor with zero thickness, for a sample with $k_s$ ranging from 0.01 to 1 W m$^{-1}$ K$^{-1}$ and a fixed $C_s$ = 0.03 MJ m$^{-3}$ K$^{-1}$. (a) Final results of the 2-step fitting (solid lines) of the numerical TPS data (open circles). (b) The goodness of fit demonstrated as the root mean square error (RMSE, black circles) and the coefficient of determination $R^2$, blue squares. (c, d) The error of the identified apparent thermal conductivity ($k_{app}$) and thermal diffusivity ($\alpha_{app}$) relative to the actual input $k_s$ and $\alpha_s$ of the sample in the numerical calculation. For such ideal sensor, the fitting shows small RMSE and $R^2$ close to 1 and the identified $k_{app}$ and $C_{app}$ are both accurate with deviation mainly from numerical noise from COMSOL simulation.

A3. Sample Size and Boundary Conditions in the Numerical Simulation and Radiation in the Time Dependent Calculation

The sample size $R_s$ and $L_s$ are set to be large enough such that further increasing the sample extents by 10% does not cause noticeable change in the result. Since the maximum value of the dimensionless time range in the identification process is decided by $\max(\tau_{cu}) = \dfrac{\sqrt{(t_{max} - t_c)\alpha_s}}{r_H} = 1$ according to the standard,[9] the TPS thermal penetration depth $d \approx 2\sqrt{(t_{max} - t_c)\alpha_s} = 2r_H$ in the numerical simulation remains constant if $r_H$ is fixed and total time of the study is long enough, regardless of the variation of other parameters. When $r_H$ is change in the simulation, we make sure that $R_s$ and $L_s$ is still large enough such that changing these sample dimensions by 10% does not change the calculation result of temperature by more than 0.1%. Thus, the condition of sample size $L_s$ = 100 mm > $d$ = 2 × (6.403) = 12.806 mm is always satisfied. Therefore, the fundamental assumption of infinite sample domain in the analytical TPS model discussed in section is maintained in all cases of numerical simulation and is ruled out in the study of systematic error.

To check if convection of the air in the sample gap play a role in determine the temperature of the sensor, we calculated the conjugated laminar flow of the air the conduction in other domains with "open boundary" boundary condition at the outer edge of the fluidic air and the gravitational force considered. The result shows the maximum velocity of the air in the gap is less than $10^{-6}$ m s$^{-1}$ and the temperature difference between a pure conduction and conjugated convection and conduction is less than 0.3%. This can be understood by estimating the Rayleigh number, i.e. the ratio of the time scale for diffusive thermal transport to the time scale for natural convective thermal transport, using $Ra = g\gamma\Delta T_{air}(h_{air})^3/\alpha_{air}\upsilon_{air}$ where $g_z$ is the gravity of the earth, $\gamma$, $\alpha_{air}$ and $\upsilon_{air}$ are the room-temperature thermal expansion coefficient, thermal diffusivity and kinematic viscosity of the air, respectively, $\Delta T_{air}$ is the temperature difference across the height of the air gap, $h_{air}$ = 60 μm, $\alpha_{air}$ $\upsilon_{air}$ Assuming a large temperature difference of $\Delta T_{air}$ = 5 K across the height of the air (much larger than the numerical simulation results which is approximately 0 K) and plug in the properties of air under 1 atm at room temperature from the literature, the $Ra$ ~$10^{-4}$ is still small due to the tiny length scale, indicating that the heat transfer by natural convection of the air is essentially negligible.[46] With the same consideration for the nano- to microscale pore of the aerogel sample, the natural convection in the porous aerogel should also be ignorable.

A4. Sensitivity Analysis of the Hot Disk TPS Temperature Response

The sensitivities for the two representative materials to TIB aerogel and stainless steel to different parameters as a function of the dimensionless time $\tau = \frac{\sqrt{t\alpha_s}}{r_H}$ are shown in Fig. A4. In both cases, the temperature change has moderate sensitivity to $k_s$, $S_{k_s}$, in the typical time window of the fitting (from $\tau_{cl} \sim 0.15$ to $\tau_{cu}$ in the range labeled with the gray vertical lines) which increases as time increases, reaching approximately -0.6 as $\tau \sim 1$. The sensitivities to the $C_s$, $S_{C_s}$, are also similar for the two materials with negative values and a peak near $\tau \sim 0.4$ to $0.6$. (The sensitivity to $\alpha_s$ is the same as that of $C_s$ with an opposite sign with a fixed $k_s$ which can be readily shown with the chain rule). This indicates that the measurement should have enough sensitivity to both $k_s$ and $C_s$ in the recommended time range and increasing $k_s$ or $C_s$ will lower the temperature of the sensor, as expected. In addition, since the $S_{C_s}$ and $S_{k_s}$ have different trends without linear relation, it is possible to fit these two parameters from the same set of data which is routinely done in the experimental hot disk TPS measurements using the commercial software.

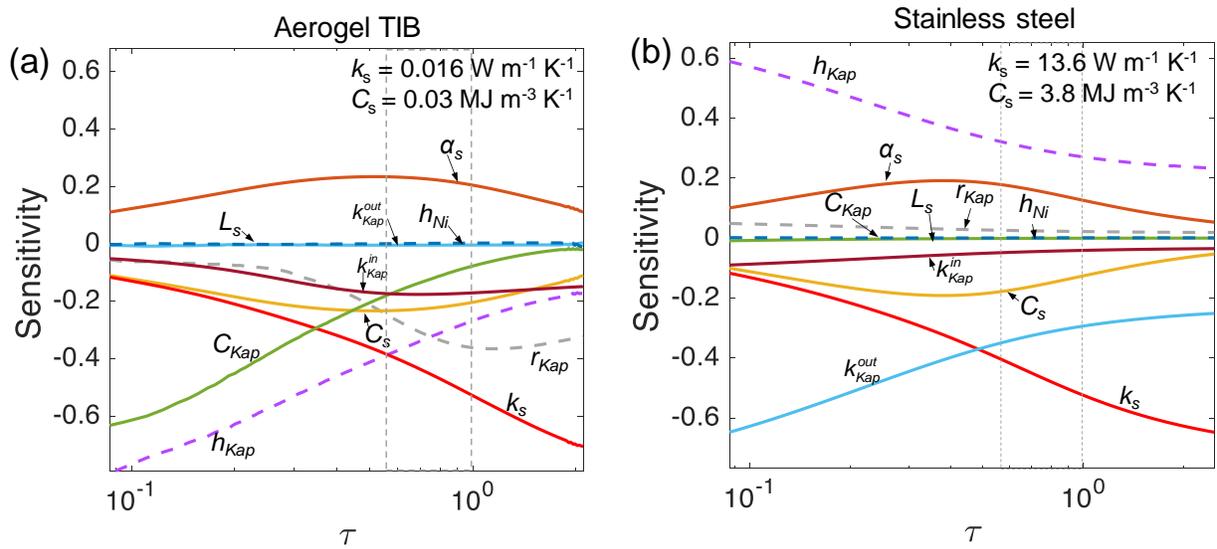

Fig. A4 Sensitivity of the average temperature rise $\Delta T$ of the Ni heater for Kapton-5501 sensor to different geometric parameters (dashed lines) and thermal properties (solid lines) for aerogel (a) at the values given in Table 1 and stainless steel (b) with all parameters held the same as Table 1 except $k_s = 13.6$ W m$^{-1}$ K$^{-1}$ and $C_s = 3.8$ MJ m$^{-3}$ K$^{-1}$. The sensitivity to each parameter is calculated by only perturbing this parameter

with all other parameters fixed at the original values except for $S_{\alpha s}$ which is simply $-S_{C_s}$ calculated with $k_s$ fixed. The Kapton radius $r_{Kap}$ is varied with the heater radius $r_H$ held constant such that this sensitivity is the sensitivity to the Kapton margin width. In both panels, the sensitivities to $L_s$ and $h_{Ni}$ are practically indistinguishable from 0. The range of the maximum dimensionless time $\tau_{cu}$ used in the identification process according to the standard,[9] $0.548 < \tau_{cu} < 1$, is indicated with thin vertical dashed lines. The dimensionless time $\tau_{cl}$ in the identification process is not specified in the standard but is required to leave at least five data points according to the commercial software.

For the aerogel sample, the sensitivities to the thickness and heat capacity of the Kapton insulation layer, $h_{Kap}$ and $C_{kap}$, are large negative numbers (hence increasing the $h_{Kap}$ and $C_{kap}$ will lower the temperature rise curve of the sensor) which decrease with time, indicating a considerable of heat diffusion and storage in the Kapton layer in a relatively short time scale $\tau < 1$. In addition, the sensitivity to both the in-plane thermal conductivity of the Kapton $k_{Kap}^{in}$ and the radius of the Kapton insulation layer $r_{Kap}$ are moderate and increase with time even at long time when the sensitivity to the $C_{kap}$ is small. (At long time, the temperature in the sensor becomes closer to isothermal state than the sample due to the high magnitude of $k_{Kap}$ compared with of the $k_{air}$ and the $k_s$.) This suggests that the evolution of temperature distribution near the sample/Kapton interface influenced by $k_{Kap}^{in}$ is important in determining the heater temperature (see also the discussion about the effective sensor radius, $r_{eff}$ in the main text). Noticeably, for such TI material, the time window for a high sensitivity to $C_{Kap}$, $r_{Kap}$, and $k_{Kap}^{in}$ largely overlap with the time window in which $S_{C_s}$ and $S_{\alpha_s}$ peak and $S_{k_s}$ is large. Such correlation makes the separation of the non-ideal sensor effect by removing early data points (through increasing $\tau_{cl}$) impossible. This kind of separation can commonly be done in the analysis for high-$k$ materials without losing measurement accuracy to $C_s$ (and $\alpha_s$) and $k_s$ as shown in the sensitivity of stainless steel in Fig. 3(b). There may also be some heat loss to the air that is in contact with the Kapton edge which is presumably less important and shall be shown in the next section.

Since the Kapton $k_{Kap}^{out}$ is one order of magnitude larger than the aerogel sample, the cross-plane thermal resistance is dominated by the sample thermal resistance in the thermal diffusion length. Since the total sensor thickness $h_{tot}$ is only 0.6% of the $r_{Kap}$ (and $k_{Kap}^{out}$ is 1/6 of $k_{Kap}^{in}$), the time scale for temperature gradient along the cross plane direction to approach the steady state is much faster than the time scale for the lateral heat diffusion and of course $\tau_{cu}$. Hence, the sensitivity to $k_{Kap}^{out}$ is negligible compared with that to $k_s$ and $k_{Kap}^{in}$. Although the sensitivities are *not* directly related to the size of error, several sources of the heat loss and heat diffusion deviating from the ideal case suggest a potentially large error in the hot disk TPS measurement of aerogel.

For the stainless steel sample, in comparison, the sensor temperature rise is sensitive to $h_{Kap}$ and $k_{Kap}^{out}$ with approximately the same magnitude but opposite signs. Meanwhile the sensitivity to the $C_{kap}$ is almost zero, suggesting a negligible fraction of heat is lost or store to the Kapton compared to the heat that diffuse into the sample. These results indicate that the Kapton layer act as a thermal resistance along the cross-plane direction similar to the interface thermal resistance (ITR) between the sensor and the sample (see the discussion for Fig. A6). Such cross-plane thermal resistance or ITR only generate a nearly constant temperature drop across the insulation layer and the interface without distorting the lateral temperature distribution in the sensor. Thus, since the linear fitting with Eq. 6 in the identification procedure to derive the slope does not care about the small vertical shift of the temperature the sensitivity of the sensor temperature response to $k_{Kap}^{out}$ does not mean it will cause error. This can also be understood from the point that the analytical model only take into account the radius of the sensor, and hence $k_{Kap}^{out}$ which has small effect on the radial temperature distribution in the sensor does not affect the identification result. The sensitivity to $k_{Kap}^{in}$, which may result from the contribution of the in-plane heat diffusion to the cross-plane thermal resistance, is small. Thus, the non-ideality of the sensor should not affect the result of the TPS measurement for stainless steel by much. This is because the thermal mass of the sensor is tiny compared to that of the steel in the volume that scales with $d$ (~$2r_H$) and the measurement time for high-k material is too short for heat to diffuse a long distance in the Kapton layer (see discussion in the next section). The small sensitivity to $r_{Kap}$ may be a result of the heat diffusion from the heated stainless steel back to the edge of the Kapton sensor since the heat flow in the steel is much faster than that in the Kapton. Based on these analyses, the hot disk TPS measurement of stainless steel should be more accurate than aerogel, consistent with experimental observations.

For both materials, the TPS sensitivity to the thickness of the Ni heater is nearly zero which is not surprising considering the fast thermalization in the high-$\alpha$ Ni layer. The negligible sensitivity to the sample dimension $L_s$ confirms our assumption that the samples size is essentially infinite.

A5. Influence of the Thermal Properties of the Insulation Layer of the Sensor and Air

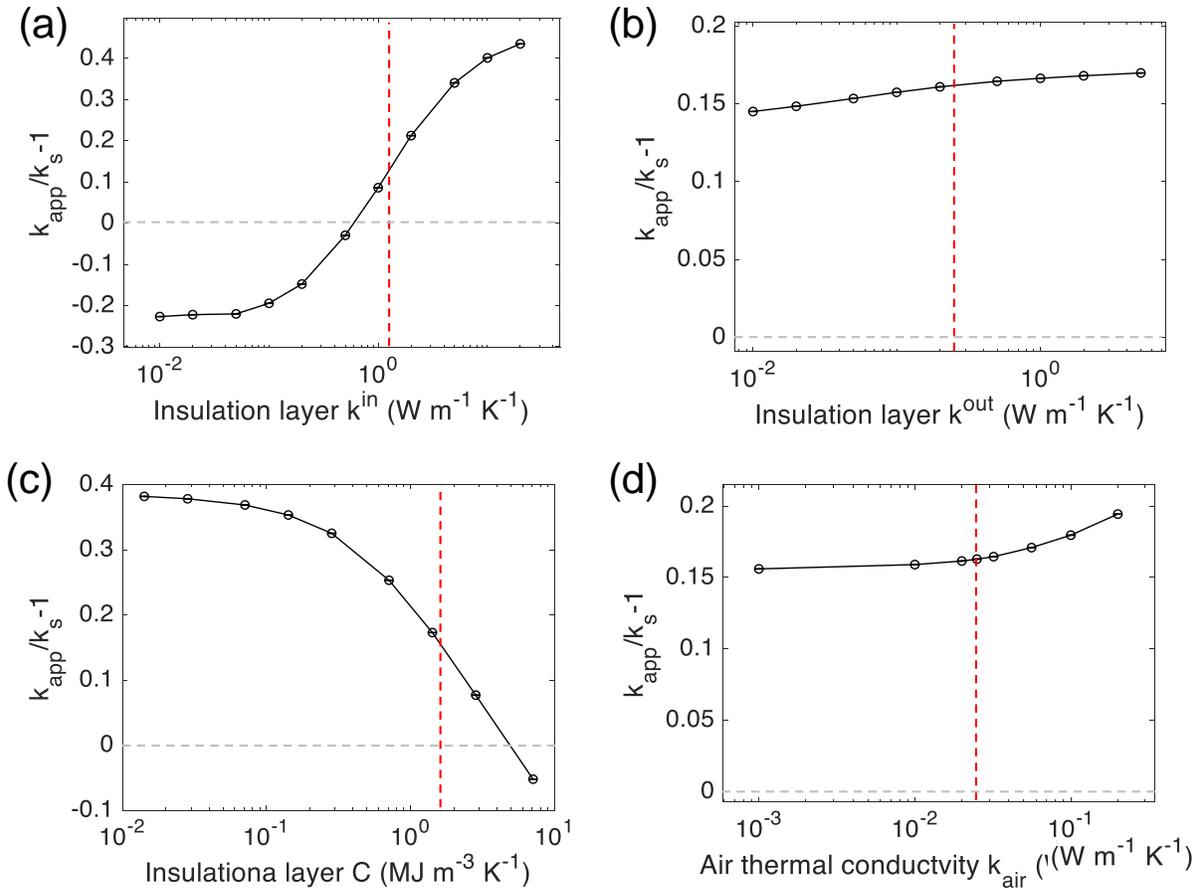

Fig. A5 Effect of the thermal properties of the insulation layer (i.e. Kapton layer in the real sensor) and air on the identified $k_{app}$ of the aerogel TIB, as compared to its true value $k_s$. The geometric parameters in Table 1 for the Kapton-5501 sensor are used and fixed. All the baseline values of the thermal properties used in the calculation are indicated by the red dashed lines. The gray dashed line labels zero along the vertical axes. When the in-plane thermal conductivity $k^{in}$ of the insulation layer is large or its C is small the in-plane heat diffusion in the insulation layer becomes important and the resulting deviation of the effective heater radius compared with $r_H$ significantly increase the overestimation, as demonstrated in (a) and (c). In comparison, the insulation layer $k_{out}$ and the corresponding out-of-plane heat diffusion (b), and conduction in air (d) have weaker influence.

In Fig. A5, we study the effect of thermal properties of the insulation layer and the air on the hot disk TPS error with a focus on the sample $k_{app}$ for only the TIB material to further identify the main source of the error. Only a single averaged absorption coefficient for the sample is considered without including the scattering in the medium. The far infrared absorption coefficient and refractive index of Kapton and silica aerogel are taken from literature at around 10 µm to approximate the actual values (not averaged over the hemisphere).[47, 48] The baseline values for the Kapton insulation layer are labeled with red dotted lines. Consistent with the sensitivity calculation and previous discussions, the heat conduction in the in-plane direction of the insulation layer has a much stronger effect than the cross-plane conduction in determining the sensor temperature and heat loss as shown in Fig. A5 (a, b). On the one hand this is because the lateral heat diffusion and storage in the insulation layer margin is related to $k^{in}$ much more than $k^{out}$ as mentioned above. On the other hand, the difference is also because of the significantly smaller $h_{Kap}$ compared to $r_{Kap}$ which causes the temperature gradient along the cross-plane direction to approach steady state much faster than the in-plane direction relative to the time window of the fitting for the TI material. A larger $k^{in}$ leads to a more significant lateral diffusion of the heat to the Kapton margin or lost to air. Increasing $k^{in}$ also increase the $r_{eff}/r_H$ which is already too large in the baseline case since $\alpha_{Kap}^{in} > \alpha_s(TIB)$. Thus, the overprediction of the sample $k_{app}$ increases with $k^{in}$ and $k^{out}$. When the $k^{in}$ is very small, $\alpha_{Kap}^{in}$ becomes smaller than $\alpha_s(TIB)$, and hence the heat loss to the edge of the Kapton is limited and the $r_{eff}$ becomes smaller than $r_H$, similar to the case of TIC in Fig. 3 in the main text, which lead to an underprediction of $k_{app}$.

As $C_{Kap}$ decreases (see Fig. A5(c)), the $\alpha_{Kap}^{in}$ increases and the temperate in the Kapton layer at the vicinity of the Kapton/sample interface increases and becomes more uniform, resulting in an increasing $r_{\text{eff}}/r_{\text{H}}$ and hence an increasing overprediction of the $k_{app}$. When $C_{\text{Kap}}$ is sufficiently small, the whole sensor acts approximately as a uniform disk heater with $r_{\text{eff}} \sim r_{\text{Kap}} > r_{\text{H}}$. As seen in Fig. A1(d), the dimensionless temperature rise of a full disk heater is close to that of a heater with relatively large ring width, hence according to Eq. A11, the temperature rise in the sample shall decreases with increasing $r_{\text{eff}}$ leading to a larger overprediction of the $k_{app}$.

Finally, as shown in Fig. A5(d), reducing the thermal conductivity of the air reduces the heat loss and the overprediction of $k_{app}$. However, even when $k_{\text{air}}$ is increased by an order of magnitude to 0.2 W m$^{-1}$ K$^{-1}$ the systematic error in $k_{app}$ still does not change by more than 5%. Thus, the heat loss to air either through the edge wall of the sensor or the interface between the sample and the air at the sensor gap is not significant in the time range of the TPS analysis. This can be understood from the fact that our simulation system has a mirror plane symmetry at $z = 0$ and the heat diffusion is strongly limited to the in-plane direction in the air which limit the heat loss.

A6. Influence of the Interfacial Thermal Resistance and the Radiative Heat Transfer

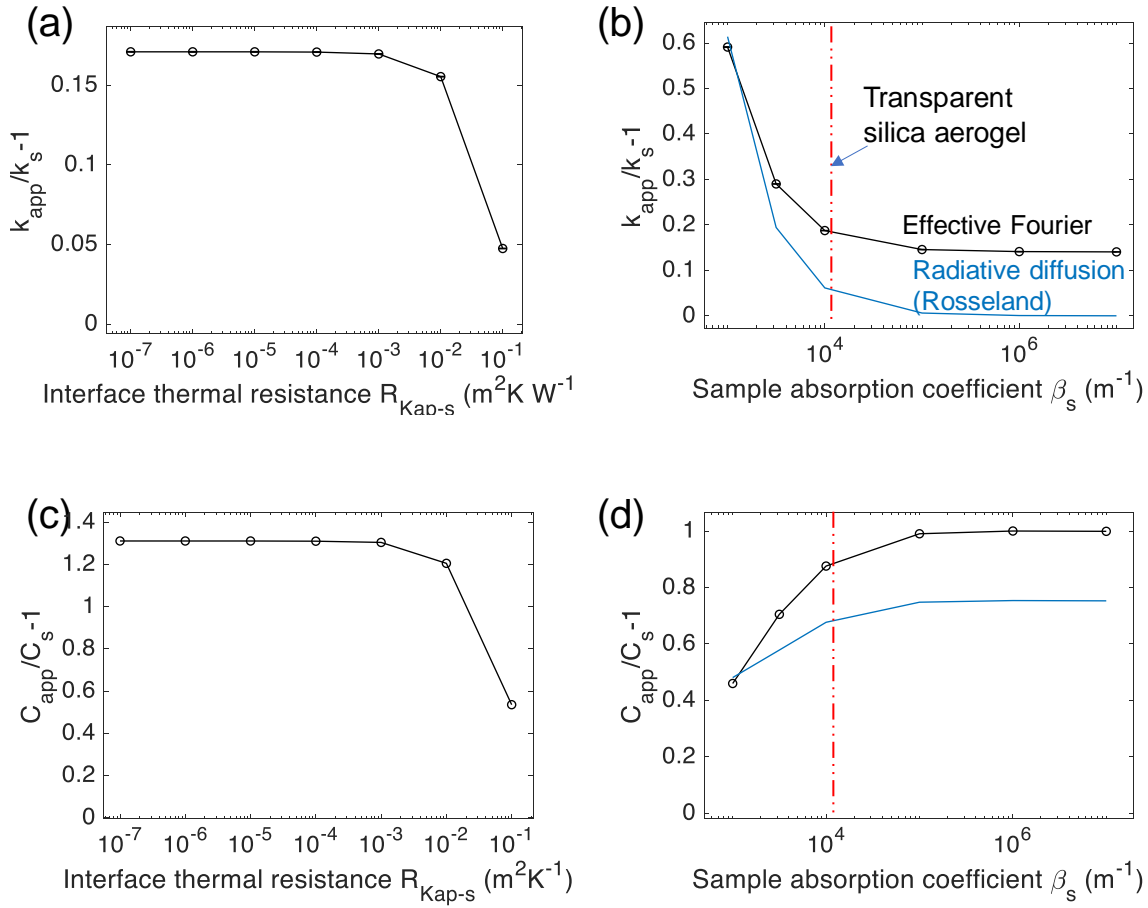

Fig. A6 Effect of thermal resistance of the interface between the Kapton insulation layer and the sample (a, c), and thermal radiation (b, d) on the identified $k_{app}$ of the aerogel TIB, as compared to its true value $k_s$. The interfacial thermal resistance (ITR) has a negligible effect on the hot disk TPS measurement when it is within the range of the typical values ($<10^{-2}\,m^2\,K^{-1}$). Panel (b) shows that the thermal radiation in a semitransparent sample contributes to the overprediction of the $k_{app}$ when the sample absorption coefficient is small. The two radiation transport models in (b, d) are explained in the main text. (Note in all other figures in this manuscript, the ITR is 0 and thermal radiation is ignored, equivalent to large $\beta_s$.) The red dashed lines in (b) and (d) label the typical absorption coefficient of silica aerogels.

In Fig. A6(a, c), we show the TPS error in the identified $k_{app}$ and $C_{app}$ with varied interfacial thermal resistance (ITR) between the Kapton layer and the sample of TIB (for both top and bottom interfaces) with all other parameters fixed at the value in Table 1. For physically relevant values in the measurement ($<10^{-2}$ m$^2$ K W$^{-1}$), changing the ITR does not affect the result of TPS measurement as it only causes approximately a constant temperature drop across the interface without largely changing the lateral heat diffusion in the sensor especially for TI materials where the measurement time is long to allow large heat diffusion length in the sample and large the thermal resistance of the sample compared to the ITR. This flat line in the error vs. ITR also confirms that the Kapton thickness dependence of the TPS error in Fig. 3(a, d) is not due to the change of the cross plane thermal resistance of the Kapton layer but rather the heat loss to the Kapton and the deviation of $r_{eff}$ from $r_H$ due to lateral heat diffusion. It is not very meaningful to talk about the drop of the error at large ITR since in that case the thermal penetration length in the sample is short and the heat is mainly trapped in the sensor. This is to some extent close to the TPS measurement of slab sample (Kapton) and the temperature response deviates a lot from the analytical bulk model though the identification process still gives a RMSE < 0.007 K for the ITR = $10^{-2}$ m$^2$ K W$^{-1}$. Changing the initial value of $\tau_{cl}$ does not help for large ITR. For non-flat sample, which is sometimes encountered in the experiment but not easy to model numerically, the contact can be bad (factor 3 of the error source in section 4.2.1) but the contact area should affect $r_{eff}$ and influence the result of the TPS measurement.

To consider the effect of radiative heat transfer on the TPS measurement accuracy, especially for semitransparent samples, we include the radiation in the participation media conjugated with the heat conduction in the COMSOL simulation. To simplify the problem, no scattering in the medium is considered and all interfaces are treated as black walls which will provide the upper limit of the radiation effect in the heat transfer problem. Sample TIB is considered with all relevant parameters listed in Table 1. The derived temperature response is analyzed in the identification process mentioned before (Fourier model) and the results of the fitted $k_{app}$ and $C_{app}$ are shown in Fig. 6 (b) and (d). Since the radiation in the sample enhance the heat transfer, the smaller the sample extinction coefficient $\beta'_s$, the larger the obtained $k_{app}$. At the same time, since radiative heat transfer is sensitive to the temperature change of the medium, it shall deviate from analytical model more at longer time of the process as sample temperature increase. Thus, as $\beta'_s$ decreases, the best linearity time range shall move $t_{max}$ to smaller value of the absolute time. Hence, with a fixed range of $\tau_{cu}$, the fitted $\alpha_{app}$ should increase and hence $C_{app}$ decreases with $\beta'_s$. When $\beta'_s$ is large, the radiative heat transfer in the sample is weak and conduction dominate; thus, the error in $k_{app}$ and $C_{app}$ saturate as $\beta'_s$ increases. However, since the radiation in the Kapton layer still enhance the heat transfer from the heater to the sensor, the heat loss is reduced and the fitted $k_{app}$ is closer to the real value compared to the previous pure conduction cases in previous Fig. 4 and 5.

Because radiation only transfers by a short distance within an optically thick medium, the radiative heat transfer process can be simplified to a diffusion process at a large scale in parallel to the conductive diffusion. The radiative heat transfer in an isotropic optically thick bulk (optical thickness $t = \sigma_{e,R}\delta \gg 1$, $\sigma_{e,R}$ is the Rosseland extinction coefficient of the medium and $\delta$ is the sample thickness), is often treated with the well-known analytical radiative diffusion approximation using the Rosseland model,[49] which predicts that the radiative thermal conductivity of optically thick medium is approximately $k_r = 16\sigma_s n^2 T^3/(3\sigma_{e,R})$, where $\sigma_s$ is the Stephen Boltzmann constant, $n$ is the medium's refractive index, $T$ is the mean medium temperature. Since the radiation effect is investigated in previous literature, we only attempt to have a qualitative understanding of the error radiation The Rosseland model results obtained by using the approximation of $\sigma_{e,R} = \beta_s$ is compared with the $k_{app}$ result identified from the COMSOL simulation data which corresponds to an effective Fourier model and shows a good agreement in the trend. Below a threshold extinction value of around $\beta_s$ ~$10^4$ m$^{-1}$, the radiative heat transfer lead to rapid increase of $k_{app}$. For silica aerogel samples, especially strengthened aerogels such TIA and TIC, a typical value $\beta_s$ is ~$10^4$-$10^5$ m$^{-1}$ which means that radiation may be responsible for less than a few percent of the overprediction of sample, considering the real interfaces are not black walls.

A6. Etching Process for Sensor Modification and the Summary of the Experimental data

For the regular Kapton-5501 sensor, a commercial Transene® Kapton Polymide Film Etchant is used for wet etching. The process is conducted at with the etchant heated to ~65 °C and the duration of each etching is 10-15 min. The Kapton-5501F sensor is dry etched with 100 W reactive-ion etching (RIE) in 200 mTorr plasma with a flow of 80 sccm $O_2$ and 20 sccm $SF_6$ for 1 h (repeated twice for the two sides of the sensor). The thickness of the sensor before and after etching is measured by a micrometer at 3-5 locations near the center of the sensor (covering the Ni heater). We confirm that after each step of etching, the total electrical resistance of the hot disk (including the sensor and the cable) obtain from the instrument is stable with a change < 3%, well within the fluctuation range of the sensor resistance due to the contact resistance as seen in pristine sensors. The same sensor is used to measure the same sample before and after both kinds of etching.

Table A1. Experimental results of all materials measured by hot disk TPS using Kapton-5501 and Kapton-5501F sensor before and after etching. The uncertainties are from standard deviation of repeated TPS measurements.

| Sensor | Stainless steel 316 | SRM1453 PS foam | Airloy x56® |
|---|---|---|---|

|  | $k_{app}$ (W m$^{-1}$ K$^{-1}$) | $C$ (MJ m$^{-3}$ K$^{-1}$) | $k_{app}$ (mW m$^{-1}$ K$^{-1}$) | $C$ (MJ m$^{-3}$ K$^{-1}$) | $k_{app}$ (mW m$^{-1}$ K$^{-1}$) | $C$ (MJ m$^{-3}$ K$^{-1}$) |
|---|---|---|---|---|---|---|
| Pristine 5501 | 13.56±0.06 | 3.77±0.09 | 33.9±0.3 | 0.027±0.0005 | 29.5±0.8 | 0.53±0.04 |
| Pristine 5501F | 13.55±0.08 | 3.91±0.08 | 34.1±0.3 | 0.030±0.001 | 29.7±0.9 | 0.52±0.05 |
| Wet etched ($h_{tot}$ = 27 µm) | 13.60±0.07 | 3.81±0.08 | 33.8±0.6 | 0.019±0.002 | 25.7±0.6 | 0.51±0.02 |
| Dry etched |  |  | 34.0±0.3* | 0.019±0.0004 | 24.2±0.6 | 0.59±0.02 |

*Data of the PS foam from the dry etched sensor is analyzed with sensor heat capacity of 4.4 mJ K$^{-1}$

Table 3. continued

| Sensor | Airloy x103® | | Hydrophobic aerogel | |
|---|---|---|---|---|
|  | $k_{app}$ (W m$^{-1}$ K$^{-1}$) | $C$ (MJ m$^{-3}$ K$^{-1}$) | $k_{app}$ (W m$^{-1}$ K$^{-1}$) | $C$ (MJ m$^{-3}$ K$^{-1}$) |
| Pristine 5501 | 31±0.5 | 0.40±0.04 | 19±1 | 0.19±0.04 |
| Pristine 5501F | 31.7±0.9 | 0.39±0.05 | 20±1 | 0.18±0.06 |
| Wet etched ($h_{tot}$ = 27 µm) | 29.2±0.8 | 0.36±0.09 | 16±1 | 0.17±0.04 |
| Dry etched | 28.0±0.6 | 0.34±0.08 | 14.8±0.8 | 0.15±0.03 |

A8. Sensor Heat Capacity in the Commercial Software

Although it is not clearly known how this is treated in the software since it is not mentioned in the standard,[9] the total heat capacity of the sensor might be treated as a lumped heat capacitor which reduce the total power output from the heater to the sample. Presumably, an iterative procedure modified based on Eq. 3-5, as mentioned in Ref [31], is used in the commercial software.

In out experiment, after both dry and wet etching, most of the Kapton on the sensor is removed and hence the total sensor heat capacity as a parameter should be reduced in the commercial software analysis though the exact amount of reduction is hard to quantify.

In Fig. 5, the experimental data for the PS foam directly from the commercial software are obtained using an appropriately reduced total sensor heat capacity linearly interpolated between the default value of $C_{\text{sensor}} = 6.4$ mJ K$^{-1}$ for the 5501F pristine sensor and a minimum $C_{\text{sensor}} = 4.4$ mJ K$^{-1}$ for the dry etched sensor due to the low $C_s$ of the sample. $C_{\text{sensor}} = 4.4$ mJ K$^{-1}$ is chosen to allow the result from the dry etched sensor to maintain the reference value of $k_{\text{app}} = 0.034$ W m$^{-1}$ K$^{-1}$. The data for aerogel sample from the dry etched sensor with the default sensor heat capacity of $C_{\text{sensor}} = 6.4$ mJ K$^{-1}$ show neglectable (<0.5%) difference with that using $C_{\text{sensor}} = 4.4$ mJ K$^{-1}$. Such relatively large $C_{\text{sensor}}$ after removing most of the Kapton layer is partially due to the large specific heat of the Ni compared with polyimide and a large volume of Ni in the whole sensor (e.g. the wide Ni lead connected to the double spiral). In addition, the model for the $C_{\text{sensor}}$ correction in the commercial software is unclear and may not capture the real heat diffusion in the sensor, thus the adjusted value $C_{\text{sensor}} = 4.4$ mJ K$^{-1}$ is not expected to match the real sensor heat capacity.

We test the dry etching result by fitting with the sensor heat capacity set to both default 6.4 mJ K$^{-1}$ and a reduced value of 4.4 mJ K$^{-1}$. The change of this parameter has negligible effect (<0.5%) on the fitting result of the high heat capacity aerogel samples including both Airloy® and the hydrophobic aerogel for both the results of the pristine sensor and etched sensor. However, for the low-$C$ PS foam sample, using the default sensor heat capacity causes the fitted $k_{\text{app}}$ to increase from the pristine result by ~10%. Only by change the sensor heat capacity to 4.4 mJ K$^{-1}$ can we reproduce the results for the PS foam. Smaller $C_{\text{sensor}}$ e.g., 3.2 mJ K$^{-1}$ will lead to $k_{\text{app}}$ of 0.0323 W m$^{-1}$ K$^{-1}$ for the PS foam sample, which is smaller than the reference value by 4%. Since the measured $k_{app}$ values for the aerogel samples using the pristine sensor with default sensor heat capacity are already overestimated in the commercial software, and the obtained $k_{app}$ is insensitive to changes in the sensor heat capacity, the overprediction problem for the aerogel samples is apparently not solvable by changing the sensor heat capacity alone. For the pristine 5501F sensor, intentionally reduce the $C_{sensor}$ from 6.4 mJ K$^{-1}$ to 0 mJ K$^{-1}$ can only change $k_{app}$ from 0.0297 W m$^{-1}$ K$^{-1}$ to 0.0294 W m$^{-1}$ K$^{-1}$ which clearly show that the method in the commercial software is not effective for TI materials.